\documentclass[11pt,a4paper]{iopart}

\usepackage{float}
\usepackage{rotating}
\usepackage{epsfig}
\usepackage[dvips]{color}

\usepackage[ansinew]{inputenc}
\usepackage[T1]{fontenc}
\usepackage[english]{babel}
\usepackage{url}
\expandafter\let\csname equation*\endcsname\relax 
\expandafter\let\csname endequation*\endcsname\relax 
\usepackage{amsmath,amssymb}
\usepackage{mathtools}
\usepackage{sistyle}
\usepackage{graphicx}

\usepackage{soul}
\sodef\spred{}{.2em}{.9em plus.4em}{1em plus.1em minus.1em}

\usepackage{tikz}
\usetikzlibrary{shapes,fadings,patterns}

\newcommand{\lp}{\left(}
\newcommand{\rp}{\right)}
\newcommand{\tb}[1]{\textbf{#1}}

\begin{document}

\title{Large--$k$ Limit of Multi--Point Propagators in the RG Formalism}
\author{Katrine Skovbo}
\address{Department of Physics and Astronomy, Aarhus University, Ny Munkegade 120, DK-8000 Aarhus C, Denmark}
\ead{kskovbo@phys.au.dk}
\date{\today}

\begin{abstract}
Renormalized versions of cosmological perturbation theory have been very successful in recent years in describing the evolution of structure formation in the weakly non--linear regime. The concept of multi--point propagators has been introduced as a tool to quantify the relation between the initial matter distribution and the final one and to push the validity of the approaches to smaller scales. We generalize the $n$--point propagators that have been considered until now to include a new class of multi--point propagators that are relevant in the framework of the renormalization group formalism. The large--$k$ results obtained for this general class of multi--point propagators match the results obtained earlier both in the case of Gaussian and non--Gaussian initial conditions. We discuss how the large--$k$ results can be used to improve on the accuracy of the calculations of the power spectrum and bispectrum in the presence of initial non--Gaussianities.
\end{abstract}

\maketitle 

\section{Introduction}\label{chap:int}

The understanding of the evolution of structures in the universe has improved dramatically with the emergence of cosmological perturbation theory in the beginning of the 1980's \cite{Juszkiewicz,Vishniac,Fry,Goroff} and the possibility of performing high resolution N--body simulations \cite{Diemand,Springel,BoylanKolchin,Heitmann}. The two approaches are complementary with perturbation theory being best suited for large scale calculations and N--body simulations working well on smaller scales. On intermediate scales the two should be compared as a consistency check on the approximations used in both approaches. The ultimate goal is to create computational approaches that can be used for sampling the cosmological parameter space and comparing with data from future galaxy surveys \cite{BOSS} and weak--lensing surveys \cite{Euclid,LSST}. This requires efficient numerical solutions and the computational cost of running N--body simulations makes it necessary to investigate semi--analytical approaches that can push the validity of perturbation theory further into the non--linear regime at small scales.

The most relevant scales for parameter estimation are those where the baryon acoustic oscillations (BAO) are most prominent at wavenumbers of $k \simeq 0.05-0.25 \; h \; \SI{}{Mpc^{-1}}$. At high redshifts standard perturbation theory describes these scales very well, but at the present time the one--loop results fail at $k \simeq 0.1 \; h \; \SI{}{Mpc^{-1}}$ and computation of the higher order corrections quickly becomes too time consuming (see \cite{Carlson} for two--loop results and \cite{Bernarint} for a thorough review of cosmological perturbation theory). New approaches are needed that either improve directly on the existing perturbation theory or formulate alternative solutions.

The similarities between cosmological perturbation theory and quantum field theory have in recent years been exploited to create a diagrammatical interpretation of the equations analogous to the Feynman diagrams of particle physics. The formalism was introduced in \cite{Scoccimarro} and has been developed in different directions in \cite{Crocce1,Valageas1,McDonald,Matarrese}. Other approaches that do not rely as heavily on understanding the field theoretical background that inspired them can be found in \cite{Pietroni,Taruya,Matsubara}. The general idea of these developments in the field is to rearrange the perturbative series and do a partial resummation of a certain class of diagrams to all orders. The validity of some of the approaches referenced here has been tested and compared against each other in \cite{Valageas3} and against N--body simulations in \cite{Carlson}. While most approaches perform better than the standard one--loop result we are still interested in pushing the validity further into the non--linear regime.

A key object in the diagrammatical formalism is the two--point propagator that describe the evolution of a Fourier mode over time. At large $k$--values, i.e. small scales where standard perturbation theory breaks down, the perturbative corrections to the two--point propagator can be resummed to gain an analytical result that is well behaved on all scales. This was first achieved in \cite{Crocce} where the result was used to define a renormalized perturbation theory and later reproduced using the renormalization group (RG) formalism in \cite{Matarrese}. In \cite{Izumi} the RG approach was generalized to deal with non--Gaussian initial conditions and a large--$k$ result was found that included a non--zero initial bispectrum.

Multi--point propagators describe a combination of time evolution and couplings between Fourier modes. The $n$--point propagators were introduced in \cite{BernardeauG} in the context of renormalized perturbation theory and it was shown that they have a similar large--$k$ behaviour as the two--point propagator. In \cite{Bernardeau} the large--$k$ behaviour was obtained also in the presence of initial non--Gaussianities. In this paper we follow the RG approach of \cite{Matarrese,Izumi} to reproduce the large--$k$ results of \cite{BernardeauG,Bernardeau} with both Gaussian and non--Gaussian initial conditions. We also generalize the concept of $n$--point propagators to include a new class of multi--point propagators that is relevant in the RG formalism and obtain large--$k$ results for these. 

In Section \ref{chap:evol} we present the equations that govern the evolution of the cosmic fields and define the multi--point propagators in the general form. Section \ref{chap:RG} contains a recap of the RG formalism with non--Gaussian initial conditions and the recipe for constructing RG equations. In Section \ref{chap:largek} we compute the large--$k$ limit of the generalized multi--point propagators and in Section \ref{chap:conc} we conclude and comment on our results.

\section{Evolution of Cosmic Fields}\label{chap:evol}

We wish to follow the evolution of matter perturbations from some initial distribution in the early universe. We use the density perturbations $\delta\lp\tb{x},\tau\rp$ defined through
\begin{align}
	\rho(\tb{x},\tau) \equiv \bar{\rho}(\tau)\lp1 + \delta(\tb{x},\tau)\rp
\end{align}
and the peculiar velocity field $\tb{v}\lp\tb{x},\tau\rp$ of the fluid. These fields satisfy the continuity and Euler equations
\begin{align}
\begin{split}
	\frac{\partial\delta}{\partial\tau} + \nabla\cdot\left[(1 + \delta)\tb{v}\right] &= 0 \\
	\frac{\partial\tb{v}}{\partial\tau} + \mathcal{H}\tb{v} + (\tb{v}\cdot\nabla)\tb{v} &= -\nabla\Psi
\end{split}
\end{align}
while the gravitational potential is sourced through the Poisson equation on subhorizon scales
\begin{align}
	\nabla^2\Psi = \frac{3}{2}\Omega_\text{m}(\tau)\mathcal{H}^2(\tau)\delta(\tb{x},\tau)
\end{align}
Here the conformal Hubble parameter is defined as $\mathcal{H} = \frac{\text{d}\log a}{\text{d}\tau} = aH$. We will assume that the background cosmology is Einstein--de Sitter with $\Omega_\text{m} = 1$ and $\Omega_\Lambda = 0$.

In the regime where our approach will be valid, i.e., large scales, we can assume that the velocity field is irrotational so that the velocity divergence, $\theta\lp\tb{x},\tau\rp \equiv \nabla\cdot\tb{v}\lp\tb{x},\tau\rp$, completely determines the velocity field. Taking the divergence of the Euler equation and going to Fourier space we obtain
\begin{align}\label{eq:Four}
\begin{split}
	\frac{\partial\delta(\tb{k},\tau)}{\partial\tau} + \theta(\tb{k},\tau) + \int{\text{d}^3\tb{q}\text{d}^3\tb{p}}\;\delta_D(\tb{k} - \tb{q} - \tb{p})\alpha(\tb{q},\tb{p})\theta(\tb{q},\tau)\delta(\tb{p},\tau) = 0& \\
	\begin{aligned}
	\frac{\partial\theta(\tb{k},\tau)}{\partial\tau} + \mathcal{H}\theta(\tb{k},\tau) &+ \frac{3}{2}\mathcal{H}^2\delta(\tb{k},\tau) \\ &+ \int{\text{d}^3\tb{q}\text{d}^3\tb{p}}\;\delta_D(\tb{k} - \tb{q} - \tb{p})\beta(\tb{q},\tb{p})\theta(\tb{q},\tau)\theta(\tb{p},\tau) = 0
	\end{aligned}&
\end{split}
\end{align}
with the two momentum factors given by
\begin{align}\label{eq:alfbet}
	\alpha(\tb{q},\tb{p}) = \frac{(\tb{q} + \tb{p})\cdot\tb{q}}{q^2} \quad \text{and} \quad \beta(\tb{q},\tb{p}) = \frac{(\tb{q} + \tb{p})^2\tb{q}\cdot\tb{p}}{2q^2p^2}
\end{align}
These factors carry the non--linearity of the equations by coupling different momentum modes of the density and velocity perturbations.

We proceed following the approach of \cite{Matarrese} and define a two--component field including the density perturbations and the velocity divergence
\begin{align} \label{eq:2cf}
	\phi_a(\tb{k},\eta) = \lp\begin{array}{c} \phi_1(\tb{k},\eta) \\ \phi_2(\tb{k},\eta) \end{array}\rp \equiv e^{-\eta} \lp\begin{array}{c} \delta(\tb{k},\eta) \\ -\theta(\tb{k},\eta)/\mathcal{H} \end{array}\rp
\end{align}
where the new time parameter is the logarithm of the scale factor $\eta = \ln(a/a_i)$ and $a_i$ corresponds to some initial time when the evolution was still in the linear regime on all relevant scales. In an Einstein--de Sitter cosmology the new time variable is equal to the logarithm of the linear growth factor, $\eta = \ln(D/D_i)$. It has been argued in \cite{Crocce} that this redefinition of $\eta$ can also be used in more general cosmologies such as $\Lambda$CDM even though it is not exact. For a scale independent linear growth factor the discrepancies affect only the decaying mode producing very small corrections to the end results at the relevant scales.

With the field $\phi_a$ the equations \eqref{eq:Four} can be combined to yield a more compact equation of motion for the fluid where repeated indices (running from 1 to 2) and momenta are being summed and integrated over respectively
\begin{align}\label{eq:eom}
	\lp\delta_{ab}\partial_\eta + \Omega_{ab}\rp\phi_b(\tb{k},\eta) = e^\eta\gamma_{abc}(\tb{k},-\tb{p},-\tb{q})\phi_b(\tb{p},\eta)\phi_c(\tb{q},\eta)
\end{align}
with
\begin{align}
	\Omega = \lp\begin{array}{cc} 1 & -1 \\ -3/2 & 3/2 \end{array}\rp
\end{align}
and the non--zero entries in the vertex factor are given by
\begin{align}\label{eq:vert}
\begin{split}
	\gamma_{121}(\tb{k},\tb{p},\tb{q}) = \gamma_{112}(\tb{k},\tb{q},\tb{p}) &= \tfrac{1}{2}\delta_D(\tb{k} + \tb{p} + \tb{q})\alpha(\tb{p},\tb{q}) \\
	\gamma_{222}(\tb{k},\tb{p},\tb{q}) &= \delta_D(\tb{k} + \tb{p} + \tb{q})\beta(\tb{p},\tb{q})
\end{split}
\end{align}
The vertex describes how two Fourier modes will interact and create structure on a different scale in the process.

In the linear regime the right hand side of equation \eqref{eq:eom} is put to zero as there is no interaction between the different modes and we can solve for the linear propagator of the system $g_{ab}$, that is the operator that evolves the linear solution $\phi_a^0$ forward in time or the Green's function of the linearized system
\begin{align}
	\phi_a^0(\tb{k},\eta_a) = g_{ab}(\eta_a,\eta_b)\phi_b^0(\tb{k},\eta_b) \quad \text{for} \quad \eta_a > \eta_b
\end{align}
By plugging this expression into equation \eqref{eq:eom} with $\gamma_{abc} = 0$ we see that the linear propagator satisfies the equation
\begin{align} \label{eq:linprop}
	\lp\delta_{ab}\partial_{\eta_a} + \Omega_{ab}\rp g_{bc}(\eta_a,\eta_b) = \delta_{ac}\delta_D(\eta_a - \eta_b)
\end{align}
which can be solved to give
\begin{align}
	g_{ab}(\eta_a,\eta_b) = \frac{1}{5}\lp \left[\begin{array}{cc} 3 & 2 \\ 3 & 2 \end{array}\right] + \left[\begin{array}{cc} 2 & -2 \\ -3 & 3 \end{array}\right]e^{-\tfrac{5}{2}(\eta_a - \eta_b)}\rp \theta_\text{H}(\eta_a - \eta_b)
\end{align}
The last matrix term with the decaying exponential factor controls the linearly decaying mode while the first term controls the linearly growing mode. By selecting appropriate initial conditions we can follow the evolution of each mode choosing $\phi_a$ proportional to $u_a = \bigl(\begin{smallmatrix} 1 \\ 1 \end{smallmatrix}\bigr)$ for the growing mode and $v_a = \bigl(\begin{smallmatrix} 1 \\ -3/2 \end{smallmatrix}\bigr)$ for the decaying mode. Due to the factor of $e^{-\eta}$ in the definition of the two--component field in equation \eqref{eq:2cf} the first term in the linear propagator reduces to a time independent constant matrix in contrast to the growing mode of the linear propagator of \cite{Crocce} that grows as $e^\eta$.

\subsection{Path Integral Formulation}

In the path integral formulation the key object is the generating functional $Z$ which can be used to generate correlation functions between the fields. It is necessary to introduce an auxiliary field $\chi_a(\textbf{k},\eta)$ that will couple to the initial conditions (see \cite{Matarrese} and \cite{Izumi} for a thorough introduction). In the case of Gaussian initial conditions the generating functional is
\begin{align} \label{eq:Zfin}
\begin{split}
	Z[J_a,K_b,P^0] =& \int{\mathcal{D}\phi_a\mathcal{D}\chi_b}\;\exp\bigg(-\frac{1}{2}\int{\text{d}\eta_a\text{d}\eta_b}\;\chi_aP^0_{ab}\delta(\eta_a)\delta(\eta_b)\chi_b \\ &+ i\int{\text{d}\eta}\,\left[\phi_ag_{ab}^{-1}\chi_b - e^\eta\gamma_{abc}\chi_a\phi_b\phi_c + J_a\phi_a + K_b\chi_b\right]\bigg)
\end{split}
\end{align}
where $P^0_{ab}$ is the initial power spectrum that completely specifies the initial Gaussian matter distribution and $J_a$ and $K_b$ are sources for the two fields $\phi_a$ and $\chi_b$ respectively. 

\begin{figure}
\begin{center}
\begin{tikzpicture}[scale=1.5,line width=1pt]
	
	\draw (-1.3cm,0) node[above] {$\eta_a$} -- (0,0);
	\draw[dashed] (0,0) -- (1.3cm,0) node[above] {$\eta_b$} node[right] {$\;\;\;\;\;\; \tb{propagator:} \; -ig_{ab}(\eta_a,\eta_b)$};
	
	\begin{scope} [yshift=-1.1cm]
	\draw[dashed] (0,0) -- (0,0.5) node[left] {$\tb{k}_a$} -- (0,0.9);
	\draw (0,0) -- (-30:0.6) node[above] {$\;\;\tb{k}_b$} -- (-30:1.1);
	\draw (0,0) -- (-150:0.6) node[above] {$\tb{k}_c\;\;\;$} -- (-150:1.1);
	\draw[white] (1.2cm,0) -- (1.3cm,0) node[right,color=black] {$\;\;\;\;\;\; \tb{vertex:} \; -ie^{\eta}\gamma_{abc}(\tb{k}_a,\tb{k}_b,\tb{k}_c)$};
	\end{scope}

	\begin{scope} [yshift=-2.1cm]
	\draw (-1.3cm,0) node[above] {$\eta_a$} -- (0,0) node[draw,rectangle,fill=white] {} -- (1.3cm,0) node[above] {$\eta_b$} node[right] {$\;\;\;\;\;\; \tb{power spectrum:} \; P^L_{ab}(k;\eta_a,\eta_b)$};
	\end{scope}
	
	\begin{scope} [yshift=-3.3cm]
	\draw (0,0) -- (0,0.5) node[left] {$\tb{k}_a$} -- (0,0.9);
	\draw (0,0) -- (-30:0.6) node[above] {$\;\;\tb{k}_b$} -- (-30:1.1);
	\draw (0,0) -- (-150:0.6) node[above] {$\tb{k}_c\;\;\;$} -- (-150:1.1);
	\draw (0,0) node[draw,rectangle,fill=white] {};
	\draw[white] (1.2cm,0) -- (1.3cm,0) node[right,color=black] {$\;\;\;\;\;\; \tb{bispectrum:} \; B^L_{abc}(\tb{k}_a,\tb{k}_b,\tb{k}_c;\eta_a,\eta_b,\eta_c)$};
	\draw[dotted] (0,-0.7cm) -- (0,-0.91cm);
	\draw[dotted] (2.5cm,-0.7cm) -- (2.5cm,-0.91cm);
	\end{scope}

\end{tikzpicture}
\end{center}
\caption{Fundamental building blocks of Feynman diagrams. The dashed lines represent the $\chi$--field while the full lines represent the $\phi$--field. Time flows right to left in the propagator and the dots represent the series of higher order statistics.}
\label{fig:fund}
\end{figure}
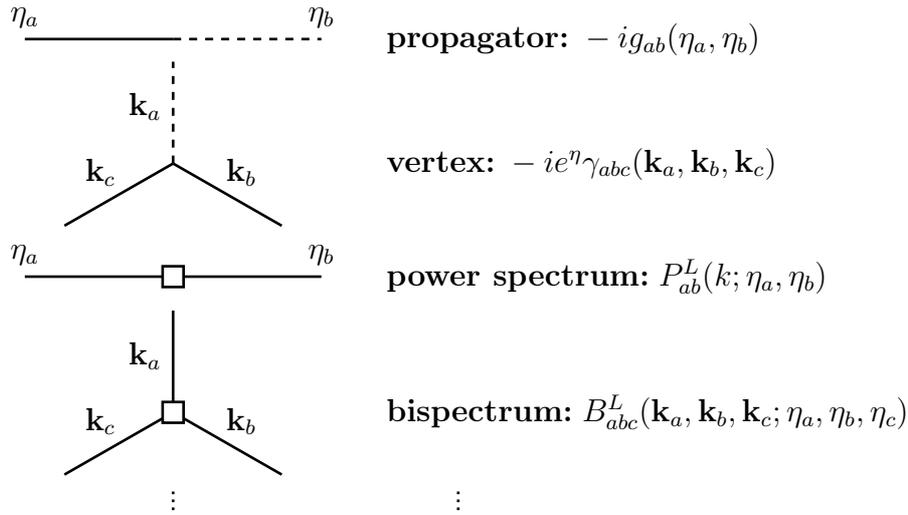

If we go back to the linear limit (setting $e^{\eta}\gamma_{abc} = 0$) the integration over the two fields can be performed explicitly yielding
\begin{align}\label{eq:Z0}
\begin{split}
	Z_0[J_a,K_b;P^0] = &\exp\Bigg(-\int{\text{d}\eta_a\text{d}\eta_b}\,\bigg[\frac{1}{2}J_a(\tb{k},\eta_a)P^L_{ab}(k;\eta_a,\eta_b)J_b(-\tb{k},\eta_b) \\ &+ iJ_a(\tb{k},\eta_a)g_{ab}(\eta_a,\eta_b)K_b(-\tb{k},\eta_b)\bigg]\Bigg)
\end{split}
\end{align}
where we see that differentiating twice with respect to $J$ gives the linearly evolved power spectrum, $P^L_{ab}(k;\eta_a,\eta_b) = g_{ac}(\eta_a,0)g_{bd}(\eta_b,0)P^0_{cd}(k)$, and differentiating once with respect to $J$ and once with respect to $K$ gives the linear propagator $g_{ab}$. These two objects along with the vertex $\gamma_{abc}$ constitute the fundamental parts of the Feynman diagrams we will consider. They will be represented by the diagrams in figure \ref{fig:fund}, where solid lines represent $\phi$ and dashed lines represent $\chi$.

In \cite{Izumi} it has been shown that in the presence of initial non--Gaussianities equations \eqref{eq:Z0} and \eqref{eq:Zfin} take the same form with equation \eqref{eq:Z0} replaced by
\begin{align}\label{eq:Z0NG}
\begin{split}
	Z_0[J_a,K_b;&P^0,B^0,\cdots] = \exp\bigg(-\frac{1}{2}\int{\text{d}\eta_a\text{d}\eta_b}\;J_a(\textbf{k},\eta_a)P^\text{L}_{ab}(k;\eta_a,\eta_b)J_b(-\textbf{k},\eta_b) \\
	&- \frac{i}{6}\int{\text{d}\eta_a\text{d}\eta_b\text{d}\eta_c}\;B^\text{L}_{abc}(\tb{k}_1,\tb{k}_2,\tb{k}_3;\eta_a,\eta_b,\eta_c)J_a(\tb{k}_1,\eta_a)J_b(\tb{k}_2,\eta_b)J_c(\tb{k}_3,\eta_c) \\
	&+ [T^\text{L}] + \cdots -i\int{\text{d}\eta_a\text{d}\eta_b}\;J_a(\textbf{k},\eta_a)g_{ab}(\eta_a,\eta_b)K_b(-\textbf{k},\eta_b)\bigg)
\end{split}
\end{align}
where $B^\text{L}_{abc}$ is the linearly evolved bispectrum, $[T^\text{L}]$ represents a similar term with the linearly evolved trispectrum and the dots represent terms from higher order statistics. Equation \eqref{eq:Z0NG} shows that differentiating $Z_0$ $n$ times with respect to the source $J_a$ will give
\begin{align}
	\frac{(-i)^n}{Z_0}\left.\frac{\delta^nZ_0}{\delta J_1(\tb{k}_1,\eta_1)\cdots\delta J_n(\tb{k}_n,\eta_n)}\right|_{J_a,K_b=0} = \delta_D\lp\textstyle\sum\tb{k}_i\rp S^\text{L}_{a_1\cdots a_n}(\tb{k}_1,\cdots,\tb{k}_n;\eta_1,\cdots,\eta_n)
\end{align}
where $S^\text{L}_{a_1\cdots a_n}$ represents the $n$'th order statistic linearly evolved in time. The linearly evolved statistics will constitute additional fundamental building blocks in the Feynman diagrams and will be represented by a box with the appropriate number of legs as shown in figure \ref{fig:fund}.

In equation \eqref{eq:Zfin} we can replace the first term in the exponential function by the infinite series
\begin{align} \label{eq:init}
\begin{split}
-\frac{1}{2}&\chi_a(\tb{k},0)P^0_{ab}(k)\chi_b(\tb{k},0) - \frac{i}{6}B^0_{abc}(\tb{k}_1,\tb{k}_2,\tb{k}_3)\chi_a(\tb{k}_1,0)\chi_b(\tb{k}_2,0)\chi_c(\tb{k}_3,0) \\ +& \frac{1}{24}T^0_{abcd}(\tb{k}_1,\tb{k}_2,\tb{k}_3,\tb{k}_4)\chi_a(\tb{k}_1,0)\chi_b(\tb{k}_2,0)\chi_c(\tb{k}_3,0)\chi_d(\tb{k}_4,0) + \cdots
\end{split}
\end{align}
including the initial bispectrum $B^0_{abc}$, trispectrum $T^0_{abcd}$ and higher order statistics, all coupled to the $\chi$--field. The fact that the initial statistics are coupled directly to the $\chi$--field only tells us that the auxiliary field plays a crucial role in connecting the evolved $\phi$--fields, $\phi(\eta)$, to the initial conditions.

Having defined the fundamental building blocks of the Feynman series we can in principle build all the non--linear results by taking loop corrections into account order by order. In this work however we will follow the renormalization group (RG) approach of \cite{Matarrese} which sets up a differential equation for the full non--linear objects that include a resummation of a series of diagrams to all orders in perturbation theory.

\subsection{Generating Functionals} \label{sec:gen}

We begin by defining the generating functional of connected diagrams $W$ 
\begin{align} \label{eq:W}
	W[J_a,K_b] = -i\log Z[J_a,K_b] 
\end{align}	
and the generating functional of one--particle--irreducible (1PI) diagrams $\Gamma$, which is given by the Legendre transform of $W$
\begin{align} \label{eq:Gam}
	\Gamma[\phi_a,\chi_b] = W[J_a,K_b] - \int{\text{d}\eta \text{d}^3\tb{k}}\,(J_a\phi_a + K_b\chi_b)
\end{align}
In this equation $\phi$ and $\chi$ actually represent the expectation values of the fields, but to keep the notation simple we will suppress this detail. We only note that the physical situation with $J_a = K_b = 0$ corresponds to the expectation values $\langle\phi_a\rangle = \langle\chi_a\rangle = 0$ because we are dealing with fluctuations away from the average of the density and velocity fields. In the presence of the sources the expectation values of the fields are given by
\begin{align}\label{eq:expec}
	\phi_a[J_a,K_b] = \frac{\delta W[J_a,K_b]}{\delta J_a} \quad \text{,} \quad \chi_b[J_a,K_b] = \frac{\delta W[J_a,K_b]}{\delta K_b}
\end{align}

The 1PI diagrams correspond to full vertices with any number of legs coming in from the right and going out to the left corresponding to differentiations of $\Gamma$ with respect to $\phi$ and $\chi$ respectively.\footnote{Notice the difference in notation as compared to \cite{BernardeauG} and \cite{Bernardeau} where $\Gamma$ represents the multi--point propagators.} The derivatives are always taken at $\phi_a = \chi_a = 0$ to recover the physical situation. As discussed in \cite{Matarrese} one can realize that derivatives of $\Gamma$ with respect to $\phi_a$ only, i.e. diagrams where every external leg represents a $\phi$--field, will vanish due to a closed loop of propagators. The second derivatives of $\Gamma$ can then be written as
\begin{align} \label{eq:Lmat}
\begin{split}
	&\Gamma^{(2)}_{\phi_a\phi_b}(k;\eta_a,\eta_b) = 0 \\
	&\Gamma^{(2)}_{\phi_a\chi_b}(k;\eta_a,\eta_b) = g_{ba}^{-1}(\eta_a,\eta_b) - \Sigma_{\phi_a\chi_b}(k;\eta_a,\eta_b) \\
	&\Gamma^{(2)}_{\chi_a\phi_b}(k;\eta_a,\eta_b) = g_{ab}^{-1}(\eta_a,\eta_b) - \Sigma_{\chi_a\phi_b}(k;\eta_a,\eta_b) \\
	&\Gamma^{(2)}_{\chi_a\chi_b}(k;\eta_a,\eta_b) = iP^0_{ab}(k)\delta(\eta_a)\delta(\eta_b) + i\Phi_{ab}(k;\eta_a,\eta_b)
\end{split}
\end{align}
where the linear part can be read off of equation \eqref{eq:Zfin} and the $\Phi$ and $\Sigma$ terms are the self--energies arising from non--linear interactions. The notation $\Gamma^{(2)}_{\phi_a\phi_b}$ represents the derivatives taken at $\phi_a = \chi_b = 0$ and includes a delta function in the momenta
\begin{align}
	\delta_D(\tb{k} + \tb{k}')\Gamma^{(2)}_{\phi_a\phi_b}(k;\eta_a,\eta_b) = \left.\frac{\delta^2\Gamma[\phi_a,\chi_b]}{\delta\phi_a(\tb{k},\eta_a)\delta\phi_b(\tb{k}',\eta_b)}\right|_{\phi_a,\chi_b=0}
\end{align}

Along the same lines as in equation \eqref{eq:Lmat} the full power spectrum and propagator can be defined in terms of the second derivatives of $W$ with respect to the sources $J$ and $K$
\begin{align} \label{eq:Wmat}
\begin{split}
	W^{(2)}_{J_aJ_b}(\tb{k},\eta_a;\tb{k}',\eta_b) &= i\delta_D(\tb{k} + \tb{k}')P_{ab}(k;\eta_a,\eta_b) \\
	W^{(2)}_{J_aK_b}(\tb{k},\eta_a;\tb{k}',\eta_b) &= -\delta_D(\tb{k} + \tb{k}')G_{ab}(k;\eta_a,\eta_b) \\
	W^{(2)}_{K_aJ_b}(\tb{k},\eta_a;\tb{k}',\eta_b) &= -\delta_D(\tb{k} + \tb{k}')G_{ba}(k;\eta_a,\eta_b) \\
	W^{(2)}_{K_aK_b}(\tb{k},\eta_a;\tb{k}',\eta_b) &= 0
\end{split}	
\end{align}
The notation $W^{(2)}_{J_aJ_b}$ represents the derivatives taken at $J_a = K_b = 0$. The component $W^{(2)}_{K_aK_b}$ being zero is a consequence of the fact that the two sets of second derivatives in equations \eqref{eq:Lmat} and \eqref{eq:Wmat} constitute inverse matrices of each other. This can also be used to express the power spectrum as a sum of two contributions
\begin{align}
	P_{ab} = P^\text{I}_{ab} + P^\text{II}_{ab}
\end{align}
given by
\begin{align} \label{eq:p1p2}
\begin{split}
	P^\text{I}_{ab}(k;\eta_a,\eta_b) &= G_{ac}(k;\eta_a,0)G_{bd}(k;\eta_b,0)P^0_{cd}(k) \\
	P^\text{II}_{ab}(k;\eta_a,\eta_b) &= \int_0^{\eta_a}{ds_1}\int_0^{\eta_b}{ds_2}\,G_{ac}(k;\eta_a,s_1)G_{bd}(k;\eta_b,s_2)\Phi_{cd}(k;s_1,s_2)
\end{split}
\end{align}
where $P^\text{I}$ is the equivalent of the linearly evolved power spectrum using the full propagator, and $P^\text{II}$ represents a new contribution from the interactions between the Fourier modes. It is clear from equation \eqref{eq:p1p2} that the full propagator and the self--energy $\Phi$ are key objects in this approach. Diagrammatically we will represent the full propagator with thick lines and $\Phi$ with a circle with two external $\chi$--legs.

The full propagator, power spectrum and 1PI $n$--point functions appearing in equations \eqref{eq:Lmat} and \eqref{eq:Wmat} represent a generalization of the corresponding linear quantities in the sense that they include all loop corrections to the given quantity. A Feynman diagram drawn with the full quantities will represent an infinite sum of diagrams in standard perturbation theory. The resummed quantities have been exploited in \cite{Crocce1} and \cite{Crocce} to build a renormalized perturbation theory of cosmological structure formation.

The full bispectrum, trispectrum and higher order statistics can be obtained by taking higher order derivatives of $W$ with respect to $J$, differentiating thrice for the bispectrum, four times for the trispectrum and so on.

\subsection{Vertices} \label{sec:vert}

Similar to equation \eqref{eq:Lmat} we can identify the full vertex with derivatives of $\Gamma$
\begin{align}\label{eq:ftv}
\begin{split}
\Gamma^{(3)}_{\chi_a\phi_b\phi_c}(\textbf{k},&s_1;-\textbf{q},s_2;-\textbf{k}+\textbf{q},s_3) \\ = &-2\delta(s_1-s)\delta(s_2-s)\delta(s_3-s)e^s\gamma_{abc}(\textbf{k},-\textbf{q},-\textbf{k}+\textbf{q}) \\ &+ \text{loop corrections}
\end{split}
\end{align}
The factor of $2$ comes from the $\phi^2$ in equation \eqref{eq:Zfin} and the delta functions account for the fact that the tree level vertex is instantaneous, i.e. it has no time dependence. The full vertex will be represented by a filled circle in the Feynman diagrams so that the diagrammatic version of equation \eqref{eq:ftv} reads
\begin{equation} \label{eq:fvert}
	\begin{tikzpicture}[line width=1pt]
	
	\draw[dashed,line width=1.7pt] (0,0) -- (-1,0);
	\draw[line width=1.7pt] (0,0) -- (30:1);
	\draw[line width=1.7pt] (0,0) -- (-30:1);
	\fill[white] (0,0) circle (1mm);
  \fill[gray,path fading=east,fading transform={rotate=-45}] (0,0) circle (1mm);  
  \draw[line width=1.7pt] (0,0) circle (1mm);
	\path (0.7,0) -- (0.8,0) node[right] {$= 2$};
	
	\begin{scope} [xshift=2.7cm]
	\draw[dashed] (0,0) -- (-1,0);
	\draw (0,0) -- (30:1);
	\draw (0,0) -- (-30:1);
	\path (0.7,0) -- (0.8,0) node[right] {$+\; 2\;\Big($};
	\end{scope}
	
	\begin{scope} [xshift=5.6cm]
	\draw[dashed] (0,0) -- (-1,0);
	\foreach \angle in {30,-30} {\draw (0,0) -- (\angle:0.2);
	\draw[dashed] (\angle:0.2) -- (\angle:0.7);
	\draw (\angle:0.7) -- (\angle:1);}
	\draw (30:0.7) arc (30:0:0.7) node[draw,rectangle,fill=white,inner sep=1mm] {} arc (0:-30:0.7);
	\path (0.7,0) -- (0.9,0) node[right] {$+ \; (2\; \text{perm.}) \Big)$};
	\end{scope}
	
	\path (1,-1) -- (1.15,-1) node[right] {$+\; \lp\text{two-loop diagrams}\rp + \cdots$};
	
	\end{tikzpicture}
\end{equation}
where I have suppressed the delta functions in the first term.

As mentioned in Section \ref{sec:gen} this definition can be generalized to higher order vertices with any number of legs coming in from the right and going out to the left
\begin{equation}
	\begin{tikzpicture} [scale=1.1,line width=1.7pt]


   	\foreach \angle in {30,20,-20} {\draw (0,0) -- (\angle:1.7cm);}
    \foreach \angle in {150,160,200} {\draw[dashed] (0,0) -- (\angle:1.7cm);}  
    
    \draw[dotted,line width=1pt] (14:1cm) arc (14:0:1cm) node[right] {$\,\,n$} arc (0:-14:1cm);
    \draw[dotted,line width=1pt] (166:1cm) arc (166:180:1cm) node[left] {$m\,\,$} arc (180:194:1cm);
    
    \fill[white] (0,0) circle (1mm);
  	\fill[gray,path fading=east,fading transform={rotate=-45}] (0,0) circle (1mm);  
  	\draw (0,0) circle (1mm);

	\end{tikzpicture}
\end{equation}
with the constraint that the sum over momenta on each side must be equal and opposite, $\tb{k}=-\sum^{n}_{i=1}\tb{k}_i=\sum^{m}_{i=1}\tb{p}_i$. The vertices defined in this way express how a number of Fourier modes can interact and combine to create power on different scales than before. This is entirely a non--linear effect as the linear evolution will preserve the initial statistics.

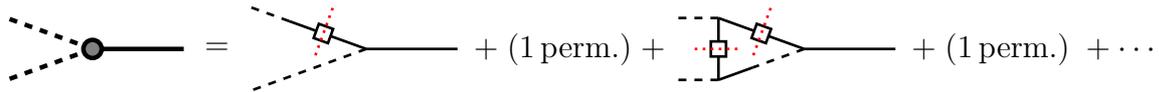
\begin{figure}
\begin{center}
	\begin{tikzpicture} [scale=1.2,line width=1pt]
	
	\begin{scope} [line width=1.7pt]
	\foreach \angle in {160,200} {\draw[dashed] (0,0) -- (\angle:1cm);}
	\draw (0,0) -- (1cm,0);
	\fill[white] (0,0) circle (1mm);
	\fill[gray,path fading=east,fading transform={rotate=-45}] (0,0) circle (1mm);  
  \draw (0,0) circle (1mm);
  \path[line width=1pt] (1cm,0) -- (1.1cm,0) node[right] {$=$};
	\end{scope}
	
	\begin{scope} [xshift=3cm]
	\draw[dashed] (0,0) -- (200:1.4cm);
	\draw (0,0) -- (160:0.5cm) node[draw,rectangle,fill=white,inner sep=1mm,rotate=-20] {} -- (160:0.9cm);
	\begin{scope} [shift={(160:0.5cm)}]
		\draw[red,dotted] (70:0.3cm) -- (250:0.3cm);
	\end{scope}
	\draw[dashed] (160:0.9cm) -- (160:1.4cm);
	\draw (0,0) -- (1cm,0);
	\path (1cm,0) -- (1.05cm,0) node[right] {$+ \; (1\, \text{perm.}) \;+$};
	\end{scope}
	
	\begin{scope} [xshift=7.8cm]
	\draw (0,0) -- (160:0.5cm) node[draw,rectangle,fill=white,inner sep=1mm,rotate=-20] {} -- (160:1cm);
	\begin{scope} [shift={(160:0.5cm)}]
		\draw[red,dotted] (70:0.3cm) -- (250:0.3cm);
	\end{scope}
	\begin{scope} [shift={(160:1cm)}]
		\draw[dashed] (0,0) -- (-0.5,0);
	\end{scope}
	\draw[dashed] (0,0) -- (200:0.6cm);
	\draw (200:0.6cm) -- (200:1cm);
	\begin{scope} [shift={(200:1cm)}]
		\draw[dashed] (0,0) -- (-0.5,0);
	\end{scope}
	\draw (160:1cm) -- (200:1cm) node[pos=0.5,draw,rectangle,fill=white,inner sep=1mm] {};
	\draw[red,dotted] (-1.2cm,0) -- (-0.7cm,0);
	\draw (0,0) -- (1cm,0);
	\path (1cm,0) -- (1.05cm,0) node[right] {$+ \; (1\, \text{perm.}) \;+ \cdots$};
	\end{scope}
	
	\end{tikzpicture}
\end{center}
\caption{Decomposing a generalized vertex into $n$--point propagators. The red dotted lines indicate that the initial statistics work as couplings between individual vertices with just one outgoing $\chi$--leg as described in the text.}
\label{fig:ex}
\end{figure}

Vertices with just one outgoing leg on the left side are the basis of the $n$--point propagators introduced by \cite{BernardeauG}. Their tree level expressions can be constructed using only the tree level vertex and the linear propagator, i.e. with no couplings to the initial conditions. On the other hand constructing diagrams with more than one outgoing leg requires the use of this kind of couplings. By considering the possible diagrams one can realize that in fact they are constructed by gluing together a number of vertices with just one outgoing leg by coupling them to the initial statistics. See figure \ref{fig:ex} for an example.

The vertices constructed in this way are not fundamental objects in the renormalized perturbation theory of \cite{BernardeauG}, but in the RG approach we will see that all the possible 1PI vertices will in principle play a role, so it is relevant to gain some insight into the general behaviour of these objects as well. Before turning to the RG approach we will clarify the connection between the formalism and notation presented here and that of \cite{BernardeauG} and \cite{Bernardeau}.

\subsection{From Vertices to Multi--Point Propagators} \label{sec:con}

To get from the 1PI diagrams in Section \ref{sec:vert} to the multi--point propagators equivalent to those in \cite{BernardeauG} we need to attach propagators to all the legs of the vertices so that the resulting diagrams describe how the initial Fourier modes at a time $\eta_b$ evolve and affect each combination of Fourier modes at a later time $\eta_a$
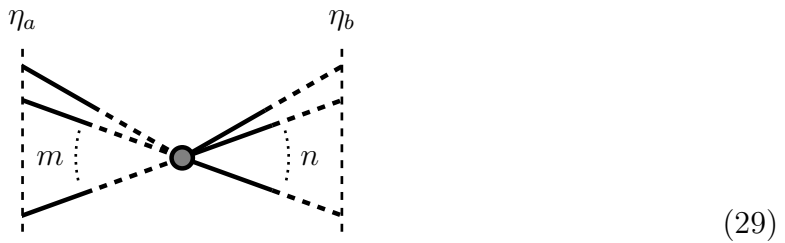
\begin{equation}
\begin{tikzpicture}[scale=1.4,line width=1.7pt]

  \begin{scope}
  \coordinate (origin) at (0,0);
  \path[clip] (-1.5cm,-0.7cm) rectangle (1.5cm,1.4cm);
  \foreach \angle in {30,20,-20} {\draw (origin) -- (\angle:0.9cm); \draw[dashed] (\angle:0.9cm) -- (\angle:2cm);}
  \foreach \angle in {150,160,200} {\draw[dashed] (origin) -- (\angle:0.9cm); \draw (\angle:0.9cm) -- (\angle:2cm);}
  \draw[dotted,line width=1pt] (15:1cm) arc (15:0:1cm) node[right] {$n$} arc (0:-15:1cm);
  \draw[dotted,line width=1pt] (165:1cm) arc (165:180:1cm) node[left] {$m$} arc (180:195:1cm);
 	\end{scope}

  \begin{scope}[line width=1pt]
  \draw[dashed] (-1.5cm,-0.7cm) -- (-1.5cm,1.1cm) node[above] {$\eta_a$}; \draw[dashed] (1.5cm,-0.7cm) -- (1.5cm,1.1cm) node[above] {$\eta_b$};
  \end{scope}
  
  \fill[white] (origin) circle (1mm);
  \fill[gray,path fading=east,fading transform={rotate=-45}] (origin) circle (1mm);  
  \draw (origin) circle (1mm);
  
\end{tikzpicture}
\end{equation}
We will denote the multi--point propagators as $V^{(n,m)}_{a_1\cdots a_m b_1\cdots b_n}$ with $n$ being the number of incoming legs and $m$ the number of outgoing legs as shown in the diagram above. As an example the formal expression for $V^{(2,1)}_{abc}$ in terms of full propagators and vertices will be
\begin{align} \label{eq:V21}
\begin{split}
V^{(2,1)}_{abc}(\tb{k}_a,\eta_a;\tb{k}_b,&\tb{k}_c,\eta_b) = \int^{\eta_a}_{\eta_b}\text{d}s_1\text{d}s_2\text{d}s_3\; G_{ad}(k_a;\eta_a,s_1) \\ \cdot&\Gamma^{(3)}_{\chi_d\phi_e\phi_f}(\tb{k}_a,s_1;\tb{k}_b,s_2;\tb{k}_c,s_3)G_{eb}(k_b;s_2,\eta_b)G_{fc}(k_c;s_3,\eta_b)
\end{split}
\end{align}
with the implicit requirement that $\tb{k}_a = -\tb{k}_b - \tb{k}_c$. In this work we will distinguish between $n$--point propagators that have $m=1$ and multi--point propagators that have $m>1$. This is in contrast to \cite{BernardeauG} and \cite{Bernardeau} where they only deal with the $m=1$ case and the two terms are used interchangeably.

The $n$--point propagator $V^{(2,1)}_{abc}$ in this work is the equivalent of the three--point propagator $\Gamma^{(2)}_{abc}$ of \cite{BernardeauG}. Apart from the differences that arise due to the extra factor of $e^{-\eta}$ in equation \eqref{eq:2cf} compared to the two--component field of \cite{BernardeauG}, the two propagators describe the same overall evolution. Naturally this equivalence holds between all the $n$--point propagators $\Gamma^{(n-1)}$ of \cite{BernardeauG} and the corresponding $V^{(n-1,1)}$ of this work. It was shown in \cite{BernardeauG} for Gaussian initial conditions and in \cite{Bernardeau} for general initial conditions that the $n$--point propagators are connected to the integration kernels of standard perturbation theory in such a way that the observable statistics at later times can be constructed directly from the $n$--point propagators. This is done in the same way as the vertices in Section \ref{sec:vert} with more than one outgoing leg are constructed by gluing together $n$--point propagators via couplings to the initial statistics. A few of the contributions to the power spectrum will be
\begin{equation}\label{eq:powser}
\begin{tikzpicture}[line width=1pt]

	\begin{scope}[line width=1.7pt]
	\draw (-0.6,0) -- (0,0) node[draw,rectangle,fill=white,inner sep=1mm] {} -- (0.6,0) node[right] {$=$};
	\end{scope}
	
	\begin{scope}[xshift=1.8cm]
	\draw (-0.6,0) -- (0,0) node[draw,rectangle,fill=white,inner sep=1mm] {} -- (0.6,0) node[right] {$+$};
	\draw[dotted,red] (0,-0.3) -- (0,0.3);
	\end{scope}
	
	\begin{scope}[xshift=4.7cm]
	\draw[dashed] (-0.6,0) -- (-1.2,0);
	\draw (-1.2,0) -- (-1.7,0);
	\draw (-0.6,0) -- (0,0.4);
	\draw (-0.6,0) -- (0,-0.4);
	\draw (0.6,0) -- (0,0.4) node[draw,rectangle,fill=white,inner sep=1mm] {};
	\draw (0.6,0) -- (0,-0.4) node[draw,rectangle,fill=white,inner sep=1mm] {};
	\draw[dashed] (0.6,0) -- (1.2,0);
	\draw (1.2,0) -- (1.7,0) node[right] {$+$};
	\draw[dotted,red] (0,-0.7) -- (0,0.7);
	\end{scope}
	
	\begin{scope}[xshift=8cm]
	\draw[dashed] (0,0) -- (-0.6,0);
	\draw (-0.6,0) -- (-1,0);
	\draw (0,0) arc (180:0:0.3);
	\draw (0,0) arc (-180:0:0.3);
	\draw (0.6,0) node[draw,rectangle,fill=white,inner sep=1mm] {} -- (1.2,0) node[right] {$+\cdots$};
	\draw[dotted,red] (0.6,0) -- (0.2,0);
	\draw[dotted,red] (0.6,0) -- (0.9,0.3);
	\draw[dotted,red] (0.6,0) -- (0.9,-0.3); 
	\end{scope}

\end{tikzpicture}
\end{equation}
where the red dotted lines again indicate where the diagrams can be split into separate $n$--point propagators. It is clear from this direct approach where all the observables can be constructed from the $n$--point propagators, that there is no need to consider the multi--point propagators with $m > 1$ in their framework. As described in Section \ref{sec:vert} this is not the case in our resummation scheme, so we will also consider the new class of multi--point propagators $V^{(n,m)}$ with $m > 1$.

We will conclude this section by studying the tree level expression for $V^{(2,1)}_{abc}$. The tree level version of equation \eqref{eq:V21} is obtained by replacing the full propagators with the linear ones and by use of the first term on the right hand side of equation \eqref{eq:ftv} for the vertex
\begin{align}\label{eq:V21tree}
	V^{(2,1)}_{abc,\,\text{tree}}(\tb{k},\eta_a;\tb{k}_1,\tb{k}_2,\eta_b) = -2\int^{\eta_a}_{\eta_b}\text{d}s\;e^sg_{ad}(\eta_a,s)\gamma_{def}(\tb{k},\tb{k}_1,\tb{k}_2)g_{eb}(s,\eta_b)g_{fc}(s,\eta_b)
\end{align}
The factor of $2$ expresses the two possible ways of contracting the vertex with the two last propagators and the overall minus sign is due to the sign convention adopted in equation \eqref{eq:Zfin}.

To compare with \cite{BernardeauG} we will compute the component with $a = 1$ and growing mode initial conditions
\begin{align}
\begin{split}
	V^{(2,1)}_{1bc,\,\text{tree}}(\tb{k},\eta_a;\tb{k}_1,\tb{k}_2,\eta_b)&u_bu_c = -2\Bigg[e^{\eta_a}\lp\frac{5}{7} + \frac{k_1x}{2k_2} + \frac{k_2x}{2k_1} + \frac{2x^2}{7}\rp \\
	- e^{\eta_b}&\lp\frac{3}{5} + \frac{k_1x}{2k_2} + \frac{k_2x}{2k_1} + \frac{2x^2}{5}\rp + e^{1/2(7\eta_b - 5\eta_a)}\lp-\frac{4}{35} + \frac{4x^2}{35}\rp\Bigg]
\end{split}
\end{align}
where $x = (\tb{k}_1\cdot\tb{k}_2)/(k_1k_2)$. We see that setting $\eta_b$ to zero and multiplying by a factor $e^{\eta_a}$ will reproduce the tree level expression of \cite{BernardeauG} apart from the overall factor of $-2$ as expected.

\section{Renormalization Group Formalism} \label{chap:RG}

The renormalization group formalism presented in \cite{Matarrese} consists of constructing differential equations for the final statistics and full propagators introduced in Sections \ref{sec:gen} and \ref{sec:con} that gradually include more and more loop corrections from the non--linearities. The resulting differential equations can in some cases be solved analytically yielding a resummation of certain loop diagrams to all orders. This is particularly useful in situations where standard perturbation theory does not produce convergent results at any finite order because the non--linear couplings are large. As can be seen in the explicit expression of $\alpha$ and $\beta$ in equation \eqref{eq:alfbet} the couplings in cosmological perturbation theory can be very large when the matter density is high. This happens at small scales (large momenta) when the initial distribution of matter has had time to collaps onto high density peaks.

\subsection{RG Equations} \label{sec:RG}

To generate the renormalization group differential equations we need to introduce a cut--off that allows only the low--order loop corrections to contribute. This requirement translates into allowing low $k$--values to contribute and can be realized by modifying the initial conditions with a step function as in \cite{Matarrese}
\begin{align} \label{eq:cut}
\begin{split}
	P^0(k) \rightarrow P^0_\lambda(k) &= P^0(k)\theta_H(\lambda - k) \\
	B^0(\tb{k}_1,\tb{k}_2,\tb{k}_3) \rightarrow B^0_\lambda(\tb{k}_1,\tb{k}_2,\tb{k}_3) &= B^0(\tb{k}_1,\tb{k}_2,\tb{k}_3)\theta_H(\lambda - (k_1 + k_2 + k_3)) \\ &\;\;\vdots
\end{split}
\end{align}
The step function, $\theta_H$, equals unity for $k<\lambda$ and zero for $k>\lambda$. For the higher order statistics there are no unique way to introduce the cut--off. Here I have used the sum of the magnitude of the wavenumbers to define the cut--off scale. Another possibility could be to use $\max\left\{k_i\right\}$. The choice of cut--off scale will affect the behaviour at low $\lambda$--values, but as we take the limit $\lambda \rightarrow \infty$ there should be no difference between the choices.

Replacing the initial statistics with the $\lambda$--dependent ones in the generating functionals and differentiating with respect to $\lambda$ will create the differential equations we are interested in. These can then be integrated from $\lambda = 0$ to $\lambda \rightarrow \infty$ to regain the full dynamics encoded in the initial conditions. We will denote the $\lambda$--dependent quantities by a lower case $\lambda$ as in equation \eqref{eq:cut}.

Because the cut--off appears only in connection with the initial statistics it is easy to calculate the $\lambda$--derivative of $Z_\lambda = Z[J_a,K_b;P^0_\lambda,B^0_\lambda,\cdots]$. Starting from the equations \eqref{eq:Zfin} and \eqref{eq:init} in the presence of initial non--Gaussianities the result is
\begin{align} \label{eq:RGZ}
\begin{split}
	\partial_\lambda Z_\lambda &= \int{\mathcal{D}\phi_a\mathcal{D}\chi_b}\,\exp(\cdots)\Bigg[-\frac{1}{2}\int{\text{d}\eta_{a,b}\text{d}^3\tb{q}}\;\delta(\eta_a)\delta(\eta_b)\chi_a\chi_b\delta(\lambda - q)P^0_{ab}(q)	\\
&-\frac{i}{6}\int{\text{d}\eta_{a,b,c}\text{d}^3\tb{q}_{1,2,3}}\;\delta(\eta_a)\delta(\eta_b)\delta(\eta_c)\chi_a\chi_b\chi_c\,\delta\big(\lambda - \textstyle\sum q_i\big) B^0_{abc}(\tb{q}_1,\tb{q}_2,\tb{q}_3) \\
	&+ \{T^0\} + \cdots\Bigg]	\\
	&= \frac{1}{2}\int{\text{d}\eta_{a,b}\text{d}^3\tb{q}}\;\delta(\lambda - q)P^0_{ab}(q)\delta(\eta_a)\delta(\eta_b)\frac{\delta^2Z_\lambda}{\delta K_a\delta K_b} \\
	&+ \frac{1}{6}\int{\text{d}\eta_{a,b,c}\text{d}^3\tb{q}_{1,2,3}}\;\delta\big(\lambda - \textstyle\sum q_i\big)B^0_{abc}(\tb{q}_1,\tb{q}_2,\tb{q}_3)\delta(\eta_a)\delta(\eta_b)\delta(\eta_c)\displaystyle\frac{\delta^3Z_\lambda}{\delta K_a\delta K_b\delta K_c} \\
	&+ \{T^0\} + \cdots
	\end{split}
\end{align}
where we have introduced the compact notation $\text{d}\eta_{a,b} = \text{d}\eta_a\text{d}\eta_b$ and $\text{d}^3\tb{q}_{1,2,3} = \text{d}^3\tb{q}_1\text{d}^3\tb{q}_2\text{d}^3\tb{q}_3$ and the argument of the exponential function in the first line can be read off of equations \eqref{eq:Zfin} and \eqref{eq:init}.

The RG equation for $W$ can be obtained by taking the $\lambda$--derivative of equation \eqref{eq:W}, $\partial_\lambda W_\lambda = -i\tfrac{1}{Z_\lambda}\partial_\lambda Z_\lambda$ and combining with equation \eqref{eq:RGZ}. The functional derivatives of $Z_\lambda$ with respect to the source field $K$ can be translated into functional derivatives of $W_\lambda$ by use of the identities
\begin{align}
\begin{split}
 \frac{\delta^2W_\lambda}{\delta K_a\delta K_b} =& -i\frac{1}{Z_\lambda}\frac{\delta^2Z_\lambda}{\delta K_a\delta K_b} + i\frac{1}{Z_\lambda}\frac{\delta Z_\lambda}{\delta K_a}\frac{1}{Z_\lambda}\frac{\delta Z_\lambda}{\delta K_b} = -i\frac{1}{Z_\lambda}\frac{\delta^2Z_\lambda}{\delta K_a\delta K_b} - i\chi_a\chi_b \\
 \frac{\delta^3W_\lambda}{\delta K_a\delta K_b\delta K_c} =& -i\frac{1}{Z_\lambda}\frac{\delta^3Z_\lambda}{\delta K_a\delta K_b\delta K_c} + \chi_a\chi_b\chi_c \\
 &- i\lp\chi_a\frac{\delta^2W_\lambda}{\delta K_b\delta K_c} + \chi_b\frac{\delta^2W_\lambda}{\delta K_a\delta K_c} + \chi_c\frac{\delta^2W_\lambda}{\delta K_a\delta K_b}\rp \\
 \vdots
\end{split}
\end{align}
where we have used equation \eqref{eq:expec} to rewrite single derivatives of $W$ with respect to $K$ in terms of the $\chi$--field. The terms that appear on the right hand side of the identities after the first term will be every combination of $\chi$--fields and functional derivatives of $W_\lambda$ of order $2$ or higher that corresponds to the number of differentiations on the left hand side. The coefficient of the first term will always be $-i$ while the coefficients of the following terms are determined by an overall factor of $i$ multiplied by an $i$ for each $\chi$ field and each $W_\lambda$ appearing in the term. The end result for the RG equation for $W_\lambda$ is
\begin{align} \label{eq:RGW}
\begin{split}
 \partial_\lambda &W_\lambda = \frac{1}{2}\int{\text{d}\eta_{a,b}\text{d}^3\tb{q}}\;\delta(\lambda - q)P^0_{ab}(q)\delta(\eta_a)\delta(\eta_b)\lp i\chi_a\chi_b + \frac{\delta^2W_\lambda}{\delta K_b\delta K_a}\rp \\
 &+ \frac{1}{6}\int{\text{d}\eta_{a,b,c}\text{d}^3\tb{q}_{1,2,3}}\;\delta\big(\lambda - \textstyle\sum q_i\big)B^0_{abc}(\tb{q}_1,\tb{q}_2,\tb{q}_3)\delta(\eta_a)\delta(\eta_b)\delta(\eta_c) \\ 
 &\times \Bigg(-\chi_a\chi_b\chi_c + i\lp\chi_a\frac{\delta^2W_\lambda}{\delta K_b\delta K_c} + \chi_b\frac{\delta^2W_\lambda}{\delta K_a\delta K_c} + \chi_c\frac{\delta^2W_\lambda}{\delta K_a\delta K_b}\rp + \frac{\delta^3W_\lambda}{\delta K_a\delta K_b\delta K_c}\Bigg) \\
 &+ \{T^0\} + \cdots
\end{split}
\end{align}
The main points to be drawn from this expression is that exactly  one initial spectrum appears in each term on the right hand side and that all the functional derivatives of $W_\lambda$ are with respect to $K$ and not $J$. This last point ensures that the functional derivatives will not generate more couplings to the initial statistics than the ones already explicitly present in equation \eqref{eq:RGW}.

From equation \eqref{eq:Gam} we see that $\partial_\lambda\Gamma_\lambda = \partial_\lambda W_\lambda$, so the RG equation for $\Gamma_\lambda$ will be similar to equation \eqref{eq:RGW} and we can use that equation to generate RG equations for any object of interest in the theory. We have not yet set the sources to zero to recover the physical situation so we keep all the terms in the parentheses on the right hand side even though they will vanish in the physical limit.

\subsection{RG Equations for Multi--Point Propagators, Gaussian Initial Conditions} \label{sec:RGpropG}

We start by considering the case of Gaussian initial conditions where only the first line of equation \eqref{eq:RGW} contributes. The two--point propagator $G_{ab}$ has already been studied using the RG formalism in \cite{Matarrese}. We will recap the results obtained there and then make the generalization to multi--point propagators.

The RG equation for the two--point propagator is obtained through the definition of $G_{ab}$ in equation \eqref{eq:Wmat}
\begin{align} \label{eq:RGprop}
\begin{split}
	\partial_\lambda G_{ab,\lambda}(k,\eta_a,\eta_b) =& -\frac{\delta^2(\partial_\lambda W_\lambda)}{\delta J_a(\tb{k},\eta_a)\delta K_b(-\tb{k},\eta_b)}\Bigg|_{J_a,K_b=0} \\ 	
	=& -\frac{1}{2}\int{\text{d}\eta_c\text{d}\eta_d\text{d}^3\textbf{q}}\;\delta(\lambda - q)P^0_{cd}(q)\delta(\eta_c)\delta(\eta_d)W^{(4)}_{J_aK_bK_cK_d,\lambda}
\end{split}
\end{align}
where the sources have now been set to zero. Performing the differentiations on $W$ the result can be written in terms of full vertices and propagators
\begin{align} \label{eq:W4}
\begin{split}
	W^{(4)}_{J_aK_bK_cK_d,\lambda}(\textbf{k},\eta_a;&-\textbf{k},\eta_b;\textbf{q},\eta_c;-\textbf{q},\eta_d) \\ 
	= \int{\text{d}s_1\cdots \text{d}s_4}\;& G_{ae,\lambda}(k;\eta_a,s_1)\,G_{fb,\lambda}(k;s_2,\eta_b)G_{gc,\lambda}(q;s_3,\eta_c)G_{hd,\lambda}(q;s_4,\eta_d) \\
		\cdot\bigg[\Gamma^{(4)}_{\chi_e\phi_f\phi_g\phi_h,\lambda}&(\textbf{k},s_1;-\textbf{k},s_2;\textbf{q},s_3;-\textbf{q},s_4) - 2\int{\text{d}s_5\text{d}s_6}\;G_{li,\lambda}(k-q;s_5,s_6) \\
	 \cdot\,\Gamma^{(3)}_{\chi_e\phi_h\phi_l,\lambda}(\textbf{k},&s_1;-\textbf{q},s_4;-\textbf{k}+\textbf{q},s_5)\Gamma^{(3)}_{\chi_i\phi_g\phi_f,\lambda}(\textbf{k}-\textbf{q},s_6;\textbf{q},s_3;-\textbf{k},s_2)\bigg]
\end{split}
\end{align}

The combination $G_{gc,\lambda}(q;s_3,0)G_{hd,\lambda}(q;s_4,0)P^0_{cd}(q)\delta(\lambda - q)$ will arise naturally as the integration kernel in all RG equations, and diagrammatically it will be represented by a crossed square with two legs so that equation \eqref{eq:RGprop} has the diagrammatical form
\begin{equation} \label{eq:RGpropdia}
\begin{tikzpicture}[line width=1.7pt]
	\draw (0,0) -- (-0.7,0) node[left] {$\displaystyle\frac{\text{d}}{\text{d}\lambda}$};
	\draw[dashed] (0,0) -- (0.8,0);
	
	\begin{scope} [xshift=3.2cm]
	\draw[dashed] (0,0) -- (-0.8,0);
	\draw (-0.8,0) -- (-1.4,0) node[left] {$= \displaystyle\frac{1}{2}$};
	\draw (0,0.6) arc (90:270:0.3);
	\draw (0,0) arc (-90:90:0.3) node[draw,rectangle,fill=white,inner sep=0pt,line width=1.5pt] {\begin{tikzpicture} [line width=1.5pt]
	\draw (-0.1,-0.1) -- (0.1,0.1); 
	\draw (-0.1,0.1) -- (0.1,-0.1); 
	\end{tikzpicture}};
	\draw (0,0) -- (0.6,0);
	\draw[dashed] (0.6,0) -- (1.4,0);
	\fill [white] (0,0) circle (1mm);
	\fill[gray,path fading=east,fading transform={rotate=-45}] (0,0) circle (1mm);
	\draw[line width=1.7pt] (0,0) circle (1mm);
	\end{scope}
	
	\begin{scope} [xshift=7.3cm]
	\draw (-0.1,0) -- (-0.5,0);
	\draw[dashed] (-1.4,0) -- (-0.5,0);
	\draw (-1.4,0) -- (-1.8,0) node[left] {$+\, \displaystyle\frac{1}{2}$};
	\draw[dashed] (-0.1,0) -- (0.5,0);
	\draw (0.5,0) -- (1,0);
	\draw[dashed] (1,0) -- (1.8,0);
	\draw (0,0.6) arc (90:0:0.6);
	\draw (-0.6,0) arc (180:90:0.6) node[draw,rectangle,fill=white,inner sep=0pt,line width=1.5pt] {\begin{tikzpicture} [line width=1.5pt]
	\draw (-0.1,-0.1) -- (0.1,0.1); 
	\draw (-0.1,0.1) -- (0.1,-0.1); 
	\end{tikzpicture}};
	\fill [white] (-0.6,0) circle (1mm);
	\fill[gray,path fading=east,fading transform={rotate=-45}] (-0.6,0) circle (1mm);
	\draw[line width=1.7pt] (-0.6,0) circle (1mm);
	\fill [white] (0.6,0) circle (1mm);
	\fill[gray,path fading=east,fading transform={rotate=-45}] (0.6,0) circle (1mm);
	\draw[line width=1.7pt] (0.6,0) circle (1mm);
	\end{scope}
	
\end{tikzpicture}
\end{equation}
The above equation represents a general rule for constructing RG equations in the case of Gaussian initial conditions, where only one--loop diagrams are present on the right hand side. This simple recipe arises from the fact mentioned in Section \ref{sec:RG} that only one initial power spectrum appears on the right hand side of equation \eqref{eq:RGW} and loops can only be constructed by use of couplings to the initial statistics.

The general rules as set up by \cite{Matarrese} for generating the right hand side of the RG equation for a given quantity in the Gaussian case are
\begin{itemize}
	\item Draw all one--loop corrections to the quantity using the full $\lambda$--dependent propagators, power spectra and vertices.
	\item Perform the $\lambda$--differentiation of the full expressions by considering only the explicit $\lambda$--dependence appearing in the theta function of $P^0_\lambda$.
\end{itemize}

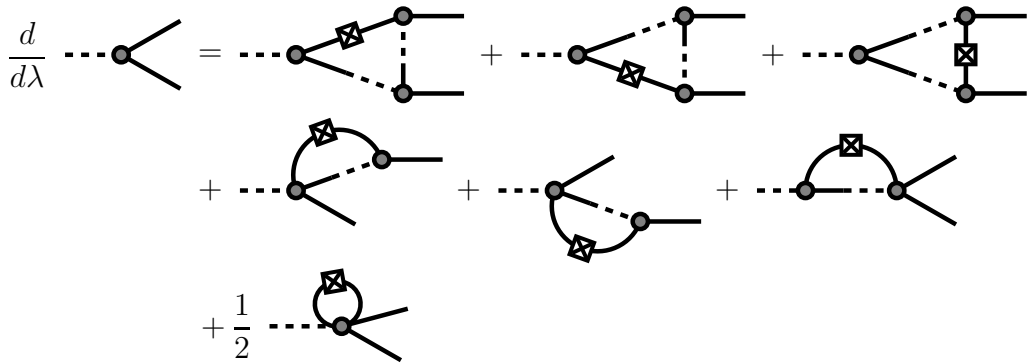
\begin{figure}
\begin{center}
\begin{tikzpicture} [line width=1.7pt]

	\draw[dashed] (0,0) -- (-0.8,0) node[left] {$\displaystyle\frac{d}{d\lambda}$};
	\draw (0,0) -- (30:0.9);
	\draw (0,0) -- (-30:0.9);
	\fill[white] (0,0) circle (1mm);
  \fill[gray,path fading=east,fading transform={rotate=-45}] (0,0) circle (1mm);  
  \draw (0,0) circle (1mm);

	\begin{scope} [xshift=2.3cm]
	\draw[dashed] (0,0) -- (-0.8,0) node[left] {$=$};
	\draw (0,0) -- (-20:0.7);
	\draw[dashed] (-20:0.7) -- (-20:1.5);
	\draw (0,0) -- (20:0.75) node[draw,rectangle,fill=white,inner sep=0,rotate=20,line width=1.5pt] {\begin{tikzpicture} [line width=1.5pt]
	\draw (-0.1,-0.1) -- (0.1,0.1); 
	\draw (-0.1,0.1) -- (0.1,-0.1); 
	\end{tikzpicture}} -- (20:1.5);
		\begin{scope} [shift={(20:1.5)}]
		\draw[dashed] (0,0) -- (0,-0.6);
		\draw (0,0) -- (0.8,0);
		\fill[white] (0,0) circle (1mm);
  	\fill[gray,path fading=east,fading transform={rotate=-45}] (0,0) circle (1mm);  
  	\draw (0,0) circle (1mm);
		\end{scope}
		\begin{scope} [shift={(-20:1.5)}]
		\draw (0,0) -- (0,0.4);
		\draw (0,0) -- (0.8,0);
		\fill[white] (0,0) circle (1mm);
  	\fill[gray,path fading=east,fading transform={rotate=-45}] (0,0) circle (1mm);  
  	\draw (0,0) circle (1mm);
		\end{scope}
	\fill[white] (0,0) circle (1mm);
  \fill[gray,path fading=east,fading transform={rotate=-45}] (0,0) circle (1mm);  
  \draw (0,0) circle (1mm);
	\end{scope}
	
	\begin{scope} [xshift=6cm]
	\draw[dashed] (0,0) -- (-0.8,0) node[left] {$+$};
	\draw (0,0) -- (20:0.7);
	\draw[dashed] (20:0.7) -- (20:1.5);
	\draw (0,0) -- (-20:0.75) node[draw,rectangle,fill=white,inner sep=0,rotate=-20,line width=1.5pt] {\begin{tikzpicture} [line width=1.5pt]
	\draw (-0.1,-0.1) -- (0.1,0.1); 
	\draw (-0.1,0.1) -- (0.1,-0.1); 
	\end{tikzpicture}} -- (-20:1.5);
		\begin{scope} [shift={(20:1.5)}]
		\draw (0,0) -- (0,-0.4);
		\draw (0,0) -- (0.8,0);
		\fill[white] (0,0) circle (1mm);
  	\fill[gray,path fading=east,fading transform={rotate=-45}] (0,0) circle (1mm);  
  	\draw (0,0) circle (1mm);
		\end{scope}
		\begin{scope} [shift={(-20:1.5)}]
		\draw[dashed] (0,0) -- (0,0.6);
		\draw (0,0) -- (0.8,0);
		\fill[white] (0,0) circle (1mm);
  	\fill[gray,path fading=east,fading transform={rotate=-45}] (0,0) circle (1mm);  
  	\draw (0,0) circle (1mm);
		\end{scope}
	\fill[white] (0,0) circle (1mm);
  \fill[gray,path fading=east,fading transform={rotate=-45}] (0,0) circle (1mm);  
  \draw (0,0) circle (1mm);
	\end{scope}
	
	\begin{scope} [xshift=9.7cm]
	\draw[dashed] (0,0) -- (-0.8,0) node[left] {$+$};
	\draw (0,0) -- (20:0.7);
	\draw[dashed] (20:0.7) -- (20:1.5);
	\draw (0,0) -- (-20:0.7);
	\draw[dashed] (-20:0.7) -- (-20:1.5);
	\draw (20:1.5) -- (-20:1.5) node[pos=0.5,draw,rectangle,fill=white,inner sep=0,line width=1.5pt] {\begin{tikzpicture} [line width=1.5pt]
	\draw (-0.1,-0.1) -- (0.1,0.1); 
	\draw (-0.1,0.1) -- (0.1,-0.1); 
	\end{tikzpicture}};
		\begin{scope} [shift={(20:1.5)}]
		\draw (0,0) -- (0.8,0);
		\fill[white] (0,0) circle (1mm);
  	\fill[gray,path fading=east,fading transform={rotate=-45}] (0,0) circle (1mm);  
  	\draw (0,0) circle (1mm);
		\end{scope}
		\begin{scope} [shift={(-20:1.5)}]
		\draw (0,0) -- (0.8,0);
		\fill[white] (0,0) circle (1mm);
  	\fill[gray,path fading=east,fading transform={rotate=-45}] (0,0) circle (1mm);  
  	\draw (0,0) circle (1mm);
		\end{scope}
	\fill[white] (0,0) circle (1mm);
  \fill[gray,path fading=east,fading transform={rotate=-45}] (0,0) circle (1mm);  
  \draw (0,0) circle (1mm);
	\end{scope}
	
	\begin{scope} [shift={(2.3,-1.8)}]
	\draw[dashed] (0,0) -- (-0.8,0) node[left] {$+$};
	\draw (0,0) -- (-30:0.9);
		\begin{scope} [shift={(20:0.6)},rotate=20]
		\draw (-0.1,0) -- (-0.6,0);
		\draw[dashed] (-0.1,0) -- (0.6,0);
		\draw (0,0.6) arc (90:0:0.6);
		\draw (-0.6,0) arc (180:90:0.6) node[draw,rectangle,fill=white,inner sep=0pt,line width=1.5pt,rotate=20] {\begin{tikzpicture} [line width=1.5pt]
	\draw (-0.1,-0.1) -- (0.1,0.1); 
	\draw (-0.1,0.1) -- (0.1,-0.1); 
	\end{tikzpicture}};
		\end{scope}
		\begin{scope} [shift={(20:1.2)}]
		\draw (0,0) -- (0.8,0);
		\fill[white] (0,0) circle (1mm);
  	\fill[gray,path fading=east,fading transform={rotate=-45}] (0,0) circle (1mm);  
  	\draw (0,0) circle (1mm);
		\end{scope}
	\fill[white] (0,0) circle (1mm);
  \fill[gray,path fading=east,fading transform={rotate=-45}] (0,0) circle (1mm);  
  \draw (0,0) circle (1mm);
	\end{scope}
	
	\begin{scope} [shift={(5.7,-1.8)}]
	\draw[dashed] (0,0) -- (-0.8,0) node[left] {$+$};
	\draw (0,0) -- (30:0.9);
		\begin{scope} [shift={(-20:0.6)},rotate=-20]
		\draw (-0.1,0) -- (-0.6,0);
		\draw[dashed] (-0.1,0) -- (0.6,0);
		\draw (0.6,0) arc (0:-90:0.6);
		\draw (-0.6,0) arc (180:270:0.6) node[draw,rectangle,fill=white,inner sep=0pt,line width=1.5pt,rotate=-20] {\begin{tikzpicture} [line width=1.5pt]
	\draw (-0.1,-0.1) -- (0.1,0.1); 
	\draw (-0.1,0.1) -- (0.1,-0.1); 
	\end{tikzpicture}};
		\end{scope}
		\begin{scope} [shift={(-20:1.2)}]
		\draw (0,0) -- (0.8,0);
		\fill[white] (0,0) circle (1mm);
  	\fill[gray,path fading=east,fading transform={rotate=-45}] (0,0) circle (1mm);  
  	\draw (0,0) circle (1mm);
		\end{scope}
	\fill[white] (0,0) circle (1mm);
  \fill[gray,path fading=east,fading transform={rotate=-45}] (0,0) circle (1mm);  
  \draw (0,0) circle (1mm);
	\end{scope}
	
	\begin{scope} [shift={(9.6,-1.8)}]
	\draw (-0.1,0) -- (-0.5,0);
	\draw[dashed] (-0.5,0) -- (-1.3,0) node[left] {$+$};
	\draw[dashed] (-0.1,0) -- (0.5,0);
	\draw (0,0.6) arc (90:0:0.6);
	\draw (-0.6,0) arc (180:90:0.6) node[draw,rectangle,fill=white,inner sep=0pt,line width=1.5pt] {\begin{tikzpicture} [line width=1.5pt]
	\draw (-0.1,-0.1) -- (0.1,0.1); 
	\draw (-0.1,0.1) -- (0.1,-0.1); 
	\end{tikzpicture}};
		\begin{scope} [xshift=0.6cm]
		\draw (0,0) -- (30:0.9);
		\draw (0,0) -- (-30:0.9);
		\end{scope}
	\fill [white] (-0.6,0) circle (1mm);
	\fill[gray,path fading=east,fading transform={rotate=-45}] (-0.6,0) circle (1mm);
	\draw[line width=1.7pt] (-0.6,0) circle (1mm);
	\fill [white] (0.6,0) circle (1mm);
	\fill[gray,path fading=east,fading transform={rotate=-45}] (0.6,0) circle (1mm);
	\draw[line width=1.7pt] (0.6,0) circle (1mm);
	\end{scope}
	
	\begin{scope} [shift={(2.9,-3.6)}]
	\draw[dashed] (0,0) -- (-1,0) node[left] {$+\,\displaystyle\frac{1}{2}$};
	\draw (0,0) -- (-30:0.9);
	\draw (0,0) -- (15:0.9);
		\begin{scope} [rotate=10]
		\draw (0,0.6) arc (90:270:0.3);
		\draw (0,0) arc (-90:90:0.3) node[draw,rectangle,fill=white,inner sep=0pt,line width=1.5pt,rotate=10] {\begin{tikzpicture} [line width=1.5pt]
		\draw (-0.1,-0.1) -- (0.1,0.1); 
		\draw (-0.1,0.1) -- (0.1,-0.1); 
		\end{tikzpicture}};
		\end{scope}
	\fill[white] (0,0) circle (1mm);
  \fill[gray,path fading=east,fading transform={rotate=-45}] (0,0) circle (1mm);  
  \draw (0,0) circle (1mm);
	\end{scope}

\end{tikzpicture}
\end{center}
\caption{RG equation for the vertex $\Gamma^{(3)}_{\chi_a\phi_b\phi_c}$.}
\label{fig:RGvert}
\end{figure}

We can use these rules to generate the RG equation for the three--point propagator $V^{(2,1)}_{abc}$. By taking the $\lambda$--derivative of equation \eqref{eq:V21} we see the general structure of the RG equations we will encounter
\begin{align}\label{eq:dlV21}
\begin{split}
	\partial_\lambda V^{(2,1)}_{abc,\lambda} = \int\text{d}&s_{1,2,3}\;\Big(\partial_\lambda\lp G_{ad,\lambda}\rp\Gamma^{(3)}_{\chi_d\phi_e\phi_f,\lambda}G_{eb,\lambda}G_{fc} + G_{ad,\lambda}\Gamma^{(3)}_{\chi_d\phi_e\phi_f,\lambda}\partial_\lambda\lp G_{eb,\lambda}\rp G_{fc,\lambda} \\ &+ G_{ad,\lambda}\Gamma^{(3)}_{\chi_d\phi_e\phi_f,\lambda}G_{eb,\lambda}\partial_\lambda\lp G_{fc,\lambda}\rp + G_{ad,\lambda}\partial_\lambda\lp\Gamma^{(3)}_{\chi_d\phi_e\phi_f,\lambda}\rp G_{eb,\lambda}G_{fc,\lambda}\Big)
	\end{split}
\end{align}
The first three terms on the right hand side represent the RG corrections to each leg as already expressed in the RG equation for the propagator in equation \eqref{eq:RGpropdia}. The last term represents the corrections that affect the vertex directly and using the above rules we can express the vertex RG equation diagrammatically as seen in figure \ref{fig:RGvert}.

This approach generalizes to higher order multi--point propagators in the sense that the RG equation can always be separated into one--loop diagrams that represent corrections to the two--point propagators in the legs and those that represent corrections to the vertex. 

The right hand side of the RG equations for vertices with more than one $\chi$--leg will in addition to the one--loop diagrams similar to those in figure \ref{fig:RGvert} contain tree level diagrams of the form
\begin{equation}\label{eq:treeRGcon}
\begin{tikzpicture}[line width=1.7pt]

	\draw (0,0) -- (0.9,0);
	\draw[dashed] (0,0) -- (220:0.8);
	\draw (0,0) -- (170:0.6) node[draw,rectangle,fill=white,inner sep=0pt,line width=1.5pt,rotate=-10] {\begin{tikzpicture} [line width=1.5pt]
		\draw (-0.1,-0.1) -- (0.1,0.1); 
		\draw (-0.1,0.1) -- (0.1,-0.1); 
		\end{tikzpicture}} -- (170:1);
	\draw[dashed] (170:1) -- (170:1.8);
	\fill[white] (0,0) circle (1mm);
  \fill[gray,path fading=east,fading transform={rotate=-45}] (0,0) circle (1mm);  
  \draw (0,0) circle (1mm);

\end{tikzpicture}
\end{equation}
Due to the delta functions $\delta(\lambda - q)$ from the integration kernel and $\delta(\tb{q} + \tb{k}_i)$ from the propagator these diagrams will only contribute in a single point $\lambda = k_i$ and will have no effect on the final solution of the differential equation. Thus we can safely ignore these deviations from the rules stated above.

RG equations for the final power spectrum, bispectrum and higher order statistics can be generated in a similar way as in equation \eqref{eq:V21}. For the power spectrum we see from equation \eqref{eq:p1p2} that the only missing piece is the RG equation for $\Phi_\lambda$ which will include diagrams like
\begin{equation} \label{eq:RGPhi}
\begin{tikzpicture}[line width=1.7pt]
	
	\draw (-0.1,0) -- (-0.6,0);
	\draw[dashed] (-0.6,0) -- (-1.4,0);
	\draw[dashed] (-0.1,0) -- (0.6,0);
	\draw[dashed] (0.6,0) -- (1.4,0);
	\path (0.6,0) -- (1.4,0.1) node[right] {\;\;and};
	\draw (0,0.6) arc (90:0:0.6);
	\draw (-0.6,0) arc (180:90:0.6) node[draw,rectangle,fill=white,inner sep=0pt,line width=1.5pt] {\begin{tikzpicture} [line width=1.5pt]
	\draw (-0.1,-0.1) -- (0.1,0.1); 
	\draw (-0.1,0.1) -- (0.1,-0.1); 
	\end{tikzpicture}};
	\fill [white] (-0.6,0) circle (1mm);
	\fill[gray,path fading=east,fading transform={rotate=-45}] (-0.6,0) circle (1mm);
	\draw[line width=1.7pt] (-0.6,0) circle (1mm);
	\fill [white] (0.6,0) circle (1mm);
	\fill[gray,path fading=east,fading transform={rotate=-45}] (0.6,0) circle (1mm);
	\draw[line width=1.7pt] (0.6,0) circle (1mm);

	\begin{scope} [xshift=3.7cm]
	\draw[dashed] (0,0) -- (-0.8,0);
	\draw (0,0.6) arc (90:270:0.3);
	\draw (0,0) arc (-90:90:0.3) node[draw,rectangle,fill=white,inner sep=0pt,line width=1.5pt] {\begin{tikzpicture} [line width=1.5pt]
	\draw (-0.1,-0.1) -- (0.1,0.1); 
	\draw (-0.1,0.1) -- (0.1,-0.1); 
	\end{tikzpicture}};
	\draw[dashed] (0,0) -- (0.8,0);
	\fill [white] (0,0) circle (1mm);
	\fill[gray,path fading=east,fading transform={rotate=-45}] (0,0) circle (1mm);
	\draw[line width=1.7pt] (0,0) circle (1mm);
	\end{scope}

\end{tikzpicture}
\end{equation}
where we see that vertices with more than one outgoing $\chi$--leg show up for the first time. To calculate the power spectrum we encounter vertices with up to two $\chi$--legs, for the bispectrum we would have vertices with up to three $\chi$--legs and so on for higher order statistics. This is why it is relevant to generalize the $n$--point propagators of \cite{BernardeauG} to multi--point propagators with more than one outgoing leg in the RG formalism.

As a final note we see that we can obtain the one--loop result from ordinary perturbation theory by keeping all quantities in the diagrams at tree level and integrate the resulting expression over $\lambda$. In this case only the first three diagrams on the right hand side of figure \ref{fig:RGvert} will contribute due to there being higher order vertices in the last four.

\subsection{RG Equations for Multi--Point Propagators, Non--Gaussian Initial Conditions} \label{sec:RGpropNG}

\begin{figure}
\begin{center}
\begin{tikzpicture} [line width=1.7pt]

	\draw[dashed] (0,0) -- (-0.6,0);
	\draw (-0.6,0) -- (-1,0);
	\draw[dashed] (-1,0) -- (-1.7,0);
	\draw (-1.7,0) -- (-2,0);
	\draw (0,0) -- (0.3,0);
	\draw[dashed] (0.3,0) -- (1,0);
	\draw (1,0) -- (1.4,0);
	\draw[dashed] (1.4,0) -- (2,0);
	\draw (180:1.05) arc (139:41:1.4);
	\draw (0,0) -- (0,0.5) node[draw,rectangle,fill=white,inner sep=0pt,line width=1.5pt] {\begin{tikzpicture}[line width=1.5pt]
	\draw (-0.1,-0.1) -- (0.1,0.1); 
	\draw (-0.1,0.1) -- (0.1,-0.1); 
	\end{tikzpicture}};
	\fill [white] (-1,0) circle (1mm);
	\fill[gray,path fading=east,fading transform={rotate=-45}] (-1,0) circle (1mm);
	\draw[line width=1.5pt] (-1,0) circle (1mm);
	\fill [white] (0,0) circle (1mm);
	\fill[gray,path fading=east,fading transform={rotate=-45}] (0,0) circle (1mm);
	\draw[line width=1.5pt] (0,0) circle (1mm);
	\fill [white] (1,0) circle (1mm);
	\fill[gray,path fading=east,fading transform={rotate=-45}] (1,0) circle (1mm);
	\draw[line width=1.5pt] (1,0) circle (1mm);
	
	\begin{scope}[xshift=4.5cm]
	\draw (-0.1,0) -- (-0.8,0);
	\draw[dashed] (-0.8,0) -- (-1.5,0);
	\draw (-1.5,0) -- (-1.8,0);
	\draw[dashed] (-0.1,0) -- (0.8,0);
	\draw (0.8,0) -- (1.2,0);
	\draw[dashed] (1.2,0) -- (1.8,0);
	\draw (-0.8,0) arc (-180:-200:0.8);
	\draw[dashed] (-200:0.8) arc (-200:-240:0.8);
	\draw (-240:0.8) to [bend left=40] (60:0.8);
	\draw (-240:0.8) to [bend right=40] (60:0.8);
	\draw (0.8,0) arc (0:60:0.8) node[draw,rectangle,fill=white,inner sep=0pt,line width=1.5pt] {\begin{tikzpicture}[line width=1.5pt]
	\draw (-0.1,-0.1) -- (0.1,0.1); 
	\draw (-0.1,0.1) -- (0.1,-0.1); 
	\end{tikzpicture}};
	\fill [white] (-0.8,0) circle (1mm);
	\fill[gray,path fading=east,fading transform={rotate=-45}] (-0.8,0) circle (1mm);
	\draw[line width=1.5pt] (-0.8,0) circle (1mm);
	\fill [white] (0.8,0) circle (1mm);
	\fill[gray,path fading=east,fading transform={rotate=-45}] (0.8,0) circle (1mm);
	\draw[line width=1.5pt] (0.8,0) circle (1mm);
	\fill [white] (-240:0.8) circle (1mm);
	\fill[gray,path fading=east,fading transform={rotate=-45}] (-240:0.8) circle (1mm);
	\draw[line width=1.5pt] (-240:0.8) circle (1mm);
	\end{scope}
	
	\begin{scope}[xshift=8.8cm]
	\draw (-0.1,0) -- (-0.8,0);
	\draw[dashed] (-0.8,0) -- (-1.5,0);
	\draw (-1.5,0) -- (-1.8,0);
	\draw[dashed] (-0.1,0) -- (0.8,0);
	\draw (0.8,0) -- (1.2,0);
	\draw[dashed] (1.2,0) -- (1.8,0);
	\draw (0.8,0) arc (0:20:0.8);
	\draw[dashed] (20:0.8) arc (20:60:0.8);
	\draw (-240:0.8) to [bend left=40] (60:0.8);
	\draw (-240:0.8) to [bend right=40] (60:0.8);
	\draw (-0.8,0) arc (-180:-240:0.8) node[draw,rectangle,fill=white,inner sep=0pt,line width=1.5pt] {\begin{tikzpicture}[line width=1.5pt]
	\draw (-0.1,-0.1) -- (0.1,0.1); 
	\draw (-0.1,0.1) -- (0.1,-0.1); 
	\end{tikzpicture}};
	\fill [white] (-0.8,0) circle (1mm);
	\fill[gray,path fading=east,fading transform={rotate=-45}] (-0.8,0) circle (1mm);
	\draw[line width=1.5pt] (-0.8,0) circle (1mm);
	\fill [white] (0.8,0) circle (1mm);
	\fill[gray,path fading=east,fading transform={rotate=-45}] (0.8,0) circle (1mm);
	\draw[line width=1.5pt] (0.8,0) circle (1mm);
	\fill [white] (60:0.8) circle (1mm);
	\fill[gray,path fading=east,fading transform={rotate=-45}] (60:0.8) circle (1mm);
	\draw[line width=1.5pt] (60:0.8) circle (1mm);
	\end{scope}
	
	\begin{scope}[shift={(-0.5cm,-1.8cm)}]
	\draw (-0.1,0) -- (-0.8,0);
	\draw[dashed] (-0.8,0) -- (-1.5,0);
	\draw (-1.5,0) -- (-1.8,0);
	\draw[dashed] (-0.1,0) -- (0.8,0);
	\draw (0.8,0) -- (1.2,0);
	\draw[dashed] (1.2,0) -- (1.8,0);
	\draw (-0.8,0) arc (-180:-270:0.8);
	\draw (-0.8,0) to [bend right=20] (90:0.8);
	\draw (0.8,0) arc (0:90:0.8) node[draw,rectangle,fill=white,inner sep=0pt,line width=1.5pt] {\begin{tikzpicture}[line width=1.5pt]
	\draw (-0.1,-0.1) -- (0.1,0.1); 
	\draw (-0.1,0.1) -- (0.1,-0.1); 
	\end{tikzpicture}};
	\fill [white] (-0.8,0) circle (1mm);
	\fill[gray,path fading=east,fading transform={rotate=-45}] (-0.8,0) circle (1mm);
	\draw[line width=1.5pt] (-0.8,0) circle (1mm);
	\fill [white] (0.8,0) circle (1mm);
	\fill[gray,path fading=east,fading transform={rotate=-45}] (0.8,0) circle (1mm);
	\draw[line width=1.5pt] (0.8,0) circle (1mm);
	\end{scope}
	
	\begin{scope}[shift={(3.8cm,-1.8cm)}]
	\draw (-0.1,0) -- (-0.8,0);
	\draw[dashed] (-0.8,0) -- (-1.5,0);
	\draw (-1.5,0) -- (-1.8,0);
	\draw[dashed] (-0.1,0) -- (0.8,0);
	\draw (0.8,0) -- (1.2,0);
	\draw[dashed] (1.2,0) -- (1.8,0);
	\draw (-0.8,0) arc (-180:-270:0.8);
	\draw (0.8,0) to [bend left=20] (90:0.8);
	\draw (0.8,0) arc (0:90:0.8) node[draw,rectangle,fill=white,inner sep=0pt,line width=1.5pt] {\begin{tikzpicture}[line width=1.5pt]
	\draw (-0.1,-0.1) -- (0.1,0.1); 
	\draw (-0.1,0.1) -- (0.1,-0.1); 
	\end{tikzpicture}};
	\fill [white] (-0.8,0) circle (1mm);
	\fill[gray,path fading=east,fading transform={rotate=-45}] (-0.8,0) circle (1mm);
	\draw[line width=1.5pt] (-0.8,0) circle (1mm);
	\fill [white] (0.8,0) circle (1mm);
	\fill[gray,path fading=east,fading transform={rotate=-45}] (0.8,0) circle (1mm);
	\draw[line width=1.5pt] (0.8,0) circle (1mm);
	\end{scope}
	
	\begin{scope}[shift={(7.3cm,-1.8cm)}]
	\draw[dashed] (0,0) -- (-0.7,0);
	\draw (-0.7,0) -- (-1,0);
	\draw (0,0) -- (0.4,0);
	\draw[dashed] (0.4,0) -- (1,0);
	\draw (0,0) -- (120:0.3);
	\draw[dashed] (120:0.3) -- (120:0.8);
	\draw (120:0.8) to [bend left=90] (0,1);
	\draw (120:0.8) to [bend right=90] (0,1);
	\draw (0,0) to [bend right=90,distance=0.5cm] (0,1) node[draw,rectangle,fill=white,inner sep=0pt,line width=1.5pt] {\begin{tikzpicture}[line width=1.5pt]
	\draw (-0.1,-0.1) -- (0.1,0.1); 
	\draw (-0.1,0.1) -- (0.1,-0.1); 
	\end{tikzpicture}};
	\fill [white] (0,0) circle (1mm);
	\fill[gray,path fading=east,fading transform={rotate=-45}] (0,0) circle (1mm);
	\draw[line width=1.5pt] (0,0) circle (1mm);
	\fill [white] (120:0.8) circle (1mm);
	\fill[gray,path fading=east,fading transform={rotate=-45}] (120:0.8) circle (1mm);
	\draw[line width=1.5pt] (120:0.8) circle (1mm);
	\end{scope}
	
	\begin{scope}[shift={(10cm,-1.8cm)}]
	\draw[dashed] (0,0) -- (-0.7,0);
	\draw (-0.7,0) -- (-1,0);
	\draw (0,0) -- (0.4,0);
	\draw[dashed] (0.4,0) -- (1,0);
	\draw (0,0) to [bend left=90,distance=0.5cm] (0,1);
	\draw (0,0) to [bend right=90,distance=0.5cm] (0,1);
	\draw (0,0) -- (0,1) node[draw,rectangle,fill=white,inner sep=0pt,line width=1.5pt] {\begin{tikzpicture}[line width=1.5pt]
	\draw (-0.1,-0.1) -- (0.1,0.1); 
	\draw (-0.1,0.1) -- (0.1,-0.1); 
	\end{tikzpicture}};
	\fill [white] (0,0) circle (1mm);
	\fill[gray,path fading=east,fading transform={rotate=-45}] (0,0) circle (1mm);
	\draw[line width=1.5pt] (0,0) circle (1mm);
	\end{scope}
	
\end{tikzpicture}
\end{center}
\caption{All loop diagrams derived from the second term of equation \eqref{eq:RGW} for the two--point propagator.}
\label{fig:biRGG}
\end{figure}
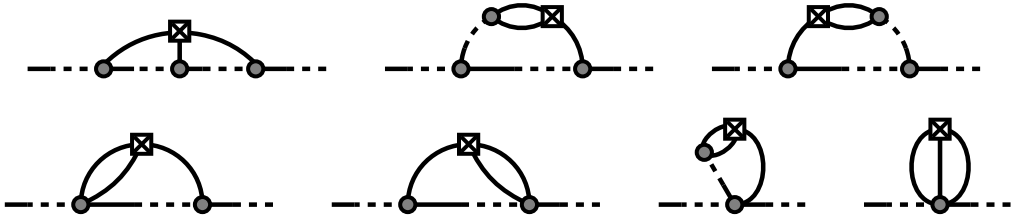

With the presence of initial non--Gaussianities the RG equations become more complicated. The first line in equation \eqref{eq:RGW} will generate the same diagrams as presented in Section \ref{sec:RGpropG}, but in addition to these the remaining terms will generate an infinite series of diagrams with up to two loops for the bispectrum, three for the trispectrum and so on. The integrations over $\tb{q}_i$ will give rise to higher order integration kernels of the form
\begin{align} \label{eq:kern} G_{ad,\lambda}(q_1;s_1,0)G_{be,\lambda}(q_2;s_2,0)G_{cf,\lambda}(q_3;s_3,0)B^0_{def}(\tb{q}_1,\tb{q}_2,\tb{q}_3)\delta\lp\lambda - \textstyle\sum q_i\rp \;\text{,}\; [T^0] \;\text{,}\;  \cdots
\end{align}
We will represent these diagrammatically as a crossed box with the appropriate number of legs for each order of the initial statistics. In accordance with the comments after equation \eqref{eq:RGW} the general rule will be that only one coupling to the initial conditions in the form of the integration kernels will be present in each diagram. Thus the general rules from Section \ref{sec:RGpropG} for constructing the right hand side of the RG equation for a quantity can be reformulated in the non--Gaussian case as
\begin{itemize}
	\item Draw all loop corrections to the quantity with exactly one crossed box using the full $\lambda$--dependent propagators,  statistics and vertices.
\end{itemize}
The differentiation with respect to $\lambda$ is already taken care of by including the crossed box. Figure \ref{fig:biRGG} shows all the diagrams arising from the bispectrum integration kernel for the two--point propagator. As can be seen all diagrams are two--loop corrections and similarly the trispectrum will only generate three--loop diagrams and so on. This will also be the case for the RG equations for vertices with just one $\chi$--leg.

For vertices with more than one $\chi$--leg there will be additional diagrams at lower order arising from each integration kernel, but the delta functions in the wavenumbers will limit the range at which they contribute. For the vertex $\Gamma^{(3)}_{\chi_a\chi_b\phi_c}$ the bispectrum will generate six additional one--loop diagrams
\begin{equation} \label{eq:bione}
\begin{tikzpicture}[line width=1.7pt]

	\draw (0,0) -- (0.5,0) node[right] {$+ \; (1\,\text{perm.}) \; +$};
	\draw[dashed] (0,0) -- (150:0.7);
	\draw (150:0.7) -- (150:1.1);
	\draw (210:1.1) -- (150:1.1);
		\begin{scope}[shift={(150:1.1)}]
		\draw[dashed] (0,0) -- (-1,0);
		\fill[white] (0,0) circle (1mm);
  	\fill[gray,path fading=east,fading transform={rotate=-45}] (0,0) circle (1mm);  
  	\draw (0,0) circle (1mm);
		\end{scope}
		\begin{scope}[shift={(210:1.1)}]
		\draw (0,0) -- (-0.4,0);
		\draw[dashed] (-0.4,0) -- (-1,0);
		\end{scope}
	\draw (0,0) -- (210:1.1) node[draw,rectangle,fill=white,inner sep=0pt,line width=1.5pt] {\begin{tikzpicture}[line width=1.5pt]
	\draw (-0.1,-0.1) -- (0.1,0.1); 
	\draw (-0.1,0.1) -- (0.1,-0.1); 
	\end{tikzpicture}};
	\fill[white] (0,0) circle (1mm);
  \fill[gray,path fading=east,fading transform={rotate=-45}] (0,0) circle (1mm);  
  \draw (0,0) circle (1mm);
	
	\begin{scope}[xshift=5cm]
	\draw (0,0) -- (0.5,0) node[right] {$+ \; (1\,\text{perm.}) \; +$};
	\draw[dashed] (0,0) -- (165:1.8);
	\draw (195:0.7) to [bend left=90,distance=3mm](195:1.3);
	\draw (195:0.7) to [bend right=90,distance=3mm](195:1.3);
	\draw (0,0) -- (195:0.7) node[draw,rectangle,fill=white,inner sep=0pt,line width=1.5pt,rotate=15] {\begin{tikzpicture}[line width=1.5pt]
	\draw (-0.1,-0.1) -- (0.1,0.1); 
	\draw (-0.1,0.1) -- (0.1,-0.1); 
	\end{tikzpicture}};
	\draw[dashed] (195:1.3) -- (195:1.9);
	\fill[white] (0,0) circle (1mm);
  \fill[gray,path fading=east,fading transform={rotate=-45}] (0,0) circle (1mm);  
  \draw (0,0) circle (1mm);
  \fill[white] (195:1.3) circle (1mm);
  \fill[gray,path fading=east,fading transform={rotate=-45}] (195:1.3) circle (1mm);  
  \draw (195:1.3) circle (1mm);
	\end{scope}
	
	\begin{scope}[xshift=9.6cm]
	\draw (0,0) -- (0.5,0) node[right] {$+ \; (1\,\text{perm.})$};
	\draw[dashed] (0,0) -- (155:1.6);
	\draw (205:0.6) -- (205:1);
	\draw[dashed] (205:1) -- (205:1.6);
	\draw (0,0) to [bend left=45,distance=3mm](205:0.6);
	\draw (0,0) to [bend right=45,distance=3mm](205:0.6) node[draw,rectangle,fill=white,inner sep=0pt,line width=1.5pt,rotate=25] {\begin{tikzpicture}[line width=1.5pt]
	\draw (-0.1,-0.1) -- (0.1,0.1); 
	\draw (-0.1,0.1) -- (0.1,-0.1); 
	\end{tikzpicture}};
	\fill[white] (0,0) circle (1mm);
  \fill[gray,path fading=east,fading transform={rotate=-45}] (0,0) circle (1mm);  
  \draw (0,0) circle (1mm);
	\end{scope}

\end{tikzpicture}
\end{equation}
When performing two of the momentum integrations in equation \eqref{eq:RGW} we get $\tb{q}_2 = -\tb{k}_1$ and $\tb{q}_3 = \tb{k}_1 - \tb{q}_1$, where $\tb{k}_1$ is the momentum associated with the lower leg to the left in the diagrams shown in equation \eqref{eq:bione}. The delta function in equation \eqref{eq:kern} will then reduce to $\delta(\lambda - q_1 - k_1 - \left|\tb{k}_1 - \tb{q}_1\right|)$, so if $k_1$ is much larger than the scale at which the bispectrum has a significant amplitude, that is $k_1 \gg q_1$, the diagrams will only contribute in a very short range $2k_1 < \lambda < 2k_1+2q_1$. The same arguments apply to loop corrections from the higher order statistics, thus in the large--$k$ limit we can neglect the contribution from the diagrams at lower order in the perturbation series arising from each initial spectrum.

It should be noted that the above simplification of the right hand side of the RG equations applies only in the large--$k$ limit. Ultimately we will be interested in calculating the final statistics at low $k$--values where the lower order diagrams cannot be neglected. In particular the power spectrum will get a new one--loop contribution from the initial bispectrum given by the last diagram in equation \eqref{eq:powser} that should be included in the calculations. We will now turn to the large--$k$ limit of the multi--point propagators and neglect these lower order diagrams.

\section{Large--$k$ Limit of Multi--Point Propagators} \label{chap:largek}

A first approximation for solving the RG equations would be to keep everything on the right hand side at its linear level. As mentioned in Section \ref{sec:RGpropG} this will reproduce the one--loop results from standard perturbation theory, so we will be interested in going beyond this level of approximations.

The large--$k$ limit of the $n$--point propagators $V^{(n-1,1)}$ has been obtained both in the Gaussian \cite{BernardeauG} and the non--Gaussian case \cite{Bernardeau}. Here we reproduce their results using the RG formalism and show that a similar large--$k$ behavior appears for the new class of multi--point propagators $V^{(n,m)} \; \text{,} \; m > 1$ introduced in Section \ref{sec:con}.

\subsection{Large--$k$ Limit with Gaussian Initial Conditions}\label{sec:largekG}

In the Gaussian case the large--$k$ regime is defined by $k_i\sigma_v
\gg 1$ where $\sigma_v^2$ is the initial velocity dispersion
\begin{align}\label{eq:sigma}
\sigma_v^2 = \frac{1}{3}\int{\text{d}^3\tb{q}}\;\frac{P^0(q)}{q^2}
\end{align}
and $k_i$ are the momenta in each leg of the multi--point propagator. In this regime the integration kernel can be kept at linear order because the dominant $q$--values are low and taking the initial fields to be in the growing mode it reduces to
\begin{equation}\label{eq:Pkerntree}
\begin{tikzpicture}[line width=1.7pt]

	\draw (-0.8,0) -- (0,0) node[draw,rectangle,fill=white,inner sep=0,line width=1.5pt] {\begin{tikzpicture} [line width=1.5pt]
	\draw (-0.1,-0.1) -- (0.1,0.1); 
	\draw (-0.1,0.1) -- (0.1,-0.1); 
	\end{tikzpicture}} -- (0.8,0) node[right] {$\approx g_{ac}(s_1,0)g_{bd}(s_2,0)u_cu_dP^0(q)\delta(\lambda - q) = u_au_b\theta_\text{H}(s_1)\theta_\text{H}(s_2)P^0(q)\delta(\lambda - q)$};
	
\end{tikzpicture}
\end{equation}
where the step functions ensure that $s_1\text{,}s_2 > 0$. We follow the approach of \cite{Matarrese}, where the large--$k$ limit of the two--point propagator has been obtained, and take the ordinary one--loop expressions for the right hand side of the RG equations as our starting point. Here we will go through a detailed calculation for the three--point propagator $V^{(2,1)}$.

\subsubsection{Large--$k$ limit of $V^{(2,1)}$, the Gaussian case}\label{sec:V21G}

Keeping the vertices at tree--level means that only the last diagram in equation \eqref{eq:RGpropdia} and the first three diagrams in figure \ref{fig:RGvert} will contribute. This gives a total of $6$ diagrams to consider, one for each propagator and three for the vertex itself. We will go through the calculations for the first vertex correction as an example. The complete diagram is
\begin{equation}
\begin{tikzpicture}[line width=1.7pt]

	\draw[dashed] (0,0) -- (-0.9,0) node[above] {$\tb{k}$};
	\draw (-0.9,0) -- (-1.5,0);
	\draw[dotted,line width=1pt] (-1.5,-0.2) -- (-1.5,0.2) node[left] {$\eta_a$};
	\draw (0,0) -- (-20:0.7);
	\draw[dashed] (-20:0.7) -- (-20:1.5);
	\draw (0,0) -- (20:0.75) node[draw,rectangle,fill=white,inner sep=0,rotate=20,line width=1.5pt] {\begin{tikzpicture} [line width=1.5pt]
	\draw (-0.1,-0.1) -- (0.1,0.1); 
	\draw (-0.1,0.1) -- (0.1,-0.1); 
	\end{tikzpicture}} -- (20:1.5);
		\begin{scope} [shift={(20:1.5)}]
		\draw[dashed] (0,0) -- (0,-0.6);
		\draw (0,0) -- (0.7,0) node[above] {$\tb{k}_1$};
		\draw[dashed] (0.7,0) -- (1.5,0);
		\fill[white] (0,0) circle (1mm);
  	\fill[gray,path fading=east,fading transform={rotate=-45}] (0,0) circle (1mm);  
  	\draw (0,0) circle (1mm);
		\end{scope}
		\begin{scope} [shift={(-20:1.5)}]
		\draw (0,0) -- (0,0.4);
		\draw (0,0) -- (0.7,0) node[above] {$\tb{k}_2$};
		\draw[dashed] (0.7,0) -- (1.5,0);
		\draw[dotted,line width=1pt] (1.5,-0.2) -- (1.5,1.3) node[right] {$\eta_b$};
		\fill[white] (0,0) circle (1mm);
  	\fill[gray,path fading=east,fading transform={rotate=-45}] (0,0) circle (1mm);  
  	\draw (0,0) circle (1mm);
		\end{scope}
	\fill[white] (0,0) circle (1mm);
  \fill[gray,path fading=east,fading transform={rotate=-45}] (0,0) circle (1mm);  
  \draw (0,0) circle (1mm);

\end{tikzpicture}
\end{equation}
Keeping everything at tree level and using equation \eqref{eq:Pkerntree} for the kernel we get
\begin{align}\label{eq:1loopcor}
\begin{split}
	-8\int{\text{d}^3\tb{q}\text{d}s_{a,b,v}}\;e^{s_a + s_b + s_v}P^0(q)\delta(\lambda -& q)\theta(s_a)\theta(s_b)g_{ad}(\eta_a,s_a)\gamma_{def}(\tb{k},-\tb{q},-\tb{k}+\tb{q})u_e \\ 
	&\times g_{fg}(s_a,s_v)\gamma_{ghi}(\tb{k}-\tb{q},\tb{k}_1+\tb{q},\tb{k}_2)g_{ic}(s_v,\eta_b) \\ 
	&\times g_{hj}(s_v,s_b)\gamma_{jkl}(-\tb{k}_1-\tb{q},\tb{q},\tb{k}_1)u_kg_{lb}(s_b,\eta_b)
	\end{split}
\end{align}
where the intermediate times are restricted by $\eta_a>s_a>s_v>s_b>\eta_b$ via the theta functions in the linear propagators.

In the $k_i \gg q$ limit two of the vertices can be reduced further using the approximation
\begin{align}\label{eq:vertred}
	\gamma_{def}(\tb{k},-\tb{q},-\tb{k}+\tb{q})u_e \approx \frac{\tb{k}\cdot\tb{q}}{2q^2}\delta_{df}
\end{align}
and the linear propagators can be contracted with these delta functions so that the dependency on $s_a$ and $s_b$ appears only in the exponential function in the first line of equation \eqref{eq:1loopcor}. Performing the two integrals over $s_a$ and $s_b$ then yields the large--$k$ result
\begin{align}\label{eq:1llk}
\begin{split}
-2\int{\text{d}s_v}\;e^{s_v}g_{ag}(\eta_a,s_v)\gamma_{ghi}(\tb{k},\tb{k}_1,\tb{k}_2)&g_{ic}(s_v,\eta_b)g_{hb}(s_v,\eta_b)(e^{\eta_a} - e^{s_v})(e^{s_v} - e^{\eta_b}) \\
&\times\int{\text{d}^3\tb{q}}\;\frac{\tb{k}\cdot\tb{q}}{q^2}\frac{\tb{k}_1\cdot\tb{q}}{q^2}P^0(q)\delta(\lambda - q)
\end{split}
\end{align}
and the momentum integration in the last line can be rewritten as
\begin{align}\label{eq:qint}
(\tb{k}\cdot\tb{k}_1)\frac{1}{3}\int{\text{d}^3\tb{q}}\;\frac{P^0(q)}{q^2}\delta(\lambda - q)
\end{align}

Going through a similar calculation for the other five diagrams will change only the momentum factor in front of the integral in equation \eqref{eq:qint} and the limits of integration over $s_a$ and $s_b$. Adding all six contributions gives
\begin{align}
\begin{split}
	\partial_\lambda V_{abc,\lambda}^{(2,1)} =& \int{\text{d}s_v}\;e^{s_v}g_{ag}(\eta_a,s_v)\gamma_{ghi}(\tb{k},\tb{k}_1,\tb{k}_2)g_{ic}(s_v,\eta_b)g_{hb}(s_v,\eta_b) \\
	&\times\Big( k^2\lp e^{\eta_a} - e^{s_v}\rp^2 + (k_1^2 + k_2^2 + 2\tb{k}_1\cdot\tb{k}_2)\lp e^{s_v} - e^{\eta_b}\rp^2 \\ 
	&- 2\tb{k}\cdot(\tb{k}_1 + \tb{k}_2)\lp e^{\eta_a} - e^{s_v}\rp\lp e^{s_v} - e^{\eta_b}\rp\Big)\frac{1}{3}\int{\text{d}^3\tb{q}}\;\frac{P^0(q)}{q^2}\delta(\lambda - q)
\end{split}
\end{align}
Using the fact that $\tb{k} = -\tb{k}_1 - \tb{k}_2$ we can reduce the sum in the parantheses to $k^2\lp e^{\eta_a} - e^{\eta_b}\rp^2$, and recognizing the tree level expression from equation \eqref{eq:V21tree} the end result for the large--$k$ limit of the one--loop calculation of the RG equation for $V^{(2,1)}$ is
\begin{align}\label{eq:V21lkdif}
\partial_\lambda V_{abc,\lambda}^{(2,1)} = - V_{abc,\,\text{tree}}^{(2,1)}k^2\frac{\lp e^{\eta_a} - e^{\eta_b}\rp^2}{2}\frac{1}{3}\int{\text{d}^3\tb{q}}\;\frac{P^0(q)}{q^2}\delta(\lambda - q)
\end{align}
We can now go beyond the usual one--loop result by promoting the tree level expression for the multi--point propagator on the right hand side to the full $\lambda$--dependent expression. The integration over $\lambda$ is then straightforward and in the limit we are interested in, $\lambda \rightarrow \infty$, the result is
\begin{align}
V_{abc,\lambda\rightarrow\infty}^{(2,1)} = V_{abc,\,\text{tree}}^{(2,1)}\exp\lp-\frac{k^2\sigma_v^2}{2}\lp e^{\eta_a} - e^{\eta_b}\rp^2\rp
\end{align}
where we have used the initial condition $V_{abc,\lambda=0}^{(2,1)} = V_{abc,\,\text{tree}}^{(2,1)}$ and $\sigma_v$ has been defined in equation \eqref{eq:sigma}. This is the result obtained in \cite{BernardeauG} for the three--point propagator by resumming an infinite series of higher order loop diagrams in standard perturbation theory. The big advantage in the RG calculation is that we need only consider the explicit expressions of one--loop diagrams.

The result shows that in the large--$k$ regime where standard perturbation theory is divergent at each order the full resummation of the dominant diagrams still gives a finite result with an exponential damping for large $k$--values.

\subsubsection{Large--$k$ limit of general multi--point propagators, the Gaussian case}

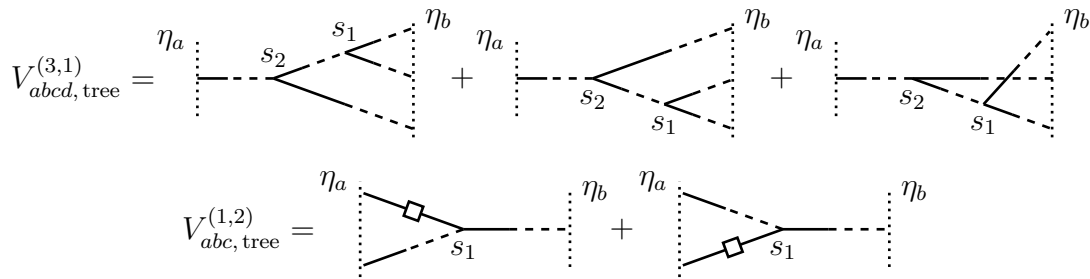
\begin{figure}
\begin{center}
\begin{tikzpicture}[line width=1pt]

	\draw[dashed] node[above] {$s_2$} (0,0) -- (-0.6,0);
	\draw (-0.6,0) -- (-1,0);
	\path (-1,0) -- (-1.4,0) node[left] {$V_{abcd,\,\text{tree}}^{(3,1)}=$};
	\draw (0,0) -- (-20:1);
	\draw[dashed] (-20:1) -- (-20:2);
	\draw (0,0) -- (20:0.4);
	\draw[dashed] (20:0.4) -- (20:1) node[above] {$s_1$};
	\draw (20:1) -- (20:1.4);
	\draw[dashed] (20:1.4) -- (20:2);
	\begin{scope}[shift={(20:1)}]
		\draw (0,0) -- (-20:0.4);
		\draw[dashed] (-20:0.4) -- (-20:1);
	\end{scope}
	\draw[dotted] (-1,-0.5) -- (-1,0.5) node[left] {$\eta_a$};
	\draw[dotted] (-23:2) -- (23:2) node[right] {$\eta_b$};

	\begin{scope}[xshift=8.4cm]
	\draw[dashed] node[below] {$s_2$} (0,0) -- (-0.6,0);
	\draw (-0.6,0) -- (-1,0);
	\path (-1,0) -- (-1.4,0) node[left] {$+$};
	\draw (0,0) -- (0.9,0);
	\draw[dashed] (0.9,0) -- (1.9,0);
	\draw (0,0) -- (-20:0.4);
	\draw[dashed] (-20:0.4) -- (-20:1) node[below] {$s_1$};
	\draw (-20:1) -- (-20:1.4);
	\draw[dashed] (-20:1.4) -- (-20:2);
	\begin{scope}[shift={(-20:1)}]
		\draw (0,0) -- (48:0.6);
		\draw[dashed] (48:0.6) -- (48:1.3);
	\end{scope}
	\draw[dotted] (-1,-0.5) -- (-1,0.5) node[left] {$\eta_a$};
	\draw[dotted] (-23:2) -- (23:2) node[right] {$\eta_b$};
	\end{scope}
	
	\begin{scope}[xshift=4.2cm]
	\draw[dashed] node[below] {$s_2$} (0,0) -- (-0.6,0);
	\draw (-0.6,0) -- (-1,0);
	\path (-1,0) -- (-1.4,0) node[left] {$+$};
	\draw (0,0) -- (20:1);
	\draw[dashed] (20:1) -- (20:2);
	\draw (0,0) -- (-20:0.4);
	\draw[dashed] (-20:0.4) -- (-20:1) node[below] {$s_1$};
	\draw (-20:1) -- (-20:1.4);
	\draw[dashed] (-20:1.4) -- (-20:2);
	\begin{scope}[shift={(-20:1)}]
		\draw (0,0) -- (20:0.4);
		\draw[dashed] (20:0.4) -- (20:1);
	\end{scope}
	\draw[dotted] (-1,-0.5) -- (-1,0.5) node[left] {$\eta_a$};
	\draw[dotted] (-23:2) -- (23:2) node[right] {$\eta_b$};
	\end{scope}
	
	\begin{scope}[shift={(2.5,-2)}]
	\path (-1,0) -- (-1.8,0) node[left] {$V_{abc,\,\text{tree}}^{(1,2)}=$};
	\draw (160:0.7) -- (160:1.4);
	\draw (0,0) -- (160:0.7) node[draw,rectangle,fill=white,inner sep=1mm,rotate=-20] {};
	\draw[dashed] (0,0) -- (200:0.8);
	\draw (200:0.8) -- (200:1.4);
	\draw node[below] {$s_1$} (0,0) -- (0.6,0);
	\draw[dashed] (0.6,0) -- (1.4,0);
	\draw[dotted] (205:1.5) -- (155:1.5) node[left] {$\eta_a$};
	\draw[dotted] (1.4,-0.5) -- (1.4,0.5) node[right] {$\eta_b$};
	\end{scope}
	
	\begin{scope}[shift={(6.7,-2)}]
	\path (-1,0) -- (-1.8,0) node[left] {$+$};
	\draw (200:0.7) -- (200:1.4);
	\draw (0,0) -- (200:0.7) node[draw,rectangle,fill=white,inner sep=1mm,rotate=20] {};
	\draw[dashed] (0,0) -- (160:0.8);
	\draw (160:0.8) -- (160:1.4);
	\draw node[below] {$s_1$} (0,0) -- (0.6,0);
	\draw[dashed] (0.6,0) -- (1.4,0);
	\draw[dotted] (205:1.5) -- (155:1.5) node[left] {$\eta_a$};
	\draw[dotted] (1.4,-0.5) -- (1.4,0.5) node[right] {$\eta_b$};
	\end{scope}

\end{tikzpicture}
\end{center}
\caption{Tree level expression for $V^{(3,1)}_{abcd}$ and $V^{(1,2)}_{abc}$.}
\label{fig:Vtree}
\end{figure}

For the general multi--point propagators there are two additional considerations that should be noted. First the tree level expressions for the diagrams will be a sum over a few different realizations that should be studied separately. Secondly the propagators with more than one outgoing $\phi$--leg will have couplings to the initial power spectrum as part of their tree level expression. See figure \ref{fig:Vtree} for two examples. We will now show that a calculation similar to that just presented can be performed for each of these tree level diagrams.

The lesson learned from the above calculation of one--loop diagrams is that in the large--$k$ regime each loop diagram can be reduced to the tree level expression multiplied by a momentum factor and a time factor as in equation \eqref{eq:1llk}. The momentum factor can be rewritten as in equation \eqref{eq:qint} where the dot product in front of the integral is determined by the two legs in the tree level diagram that the loop is attached to. The time factor arises from the integration over $e^{s_a + s_b}$ where $s_a$ and $s_b$ are the two times of interaction where the loop hits the tree level diagram. The limits of the integrations can be read off from diagrams like those in figure \ref{fig:Vtree} depending on which section of the tree level diagram the legs of the loop touch.

Let us start by considering a segment of a tree level diagram with a power spectrum. There will be three types of loop diagrams with both legs attached to such a segment
\begin{equation}
\begin{tikzpicture}[line width=1pt]

	\draw (-0.1,0) -- (-0.5,0);
	\draw[dashed] (-0.5,0) -- (-1.1,0);
	\draw (-1.1,0) -- (-1.5,0) node[below] {$s_2$};
	\draw[dashed] (-0.1,0) -- (0.5,0);
	\draw (1,0) -- (1.5,0) node[below] {$s_1$};
	\draw (0.5,0) -- (1,0) node[draw,rectangle,inner sep=1mm,fill=white] {};
	\path (1,0) -- (1,-0.3) node {$0$};
	\draw (0,0.5) arc (90:0:0.5);
	\draw (-0.5,0) arc (180:90:0.5) node[draw,rectangle,fill=white,inner sep=0pt] {\begin{tikzpicture}[line width=1pt]
	\draw (-0.1,-0.1) -- (0.1,0.1); 
	\draw (-0.1,0.1) -- (0.1,-0.1); 
	\end{tikzpicture}};
	
	\begin{scope}[xshift=3.7cm]
	\draw (0,0) -- (-0.5,0);
	\draw[dashed] (-0.5,0) -- (-1.1,0);
	\draw (-1.1,0) -- (-1.5,0) node[below] {$s_2$} node[left] {$+$};
	\draw (0.5,0) -- (0,0) node[draw,rectangle,inner sep=1mm,fill=white] {};
	\path (0,0) -- (0,-0.3) node {$0$};
	\draw[dashed] (0.5,0) -- (1.1,0);
	\draw (1.1,0) -- (1.5,0) node[below] {$s_1$};
	\draw (0,0.5) arc (90:0:0.5);
	\draw (-0.5,0) arc (180:90:0.5) node[draw,rectangle,fill=white,inner sep=0pt] {\begin{tikzpicture}[line width=1pt]
	\draw (-0.1,-0.1) -- (0.1,0.1); 
	\draw (-0.1,0.1) -- (0.1,-0.1); 
	\end{tikzpicture}};
	\end{scope}

	\begin{scope}[xshift=7.4cm]
	\draw[dashed] (0.1,0) -- (-0.5,0);
	\draw (-1,0) -- (-1.5,0) node[below] {$s_2$} node[left] {$+$};
	\draw (-0.5,0) -- (-1,0) node[draw,rectangle,inner sep=1mm,fill=white] {};
	\path (-1,0) -- (-1,-0.3) node {$0$};
	\draw (0.1,0) -- (0.5,0) ;
	\draw[dashed] (0.5,0) -- (1.1,0);
	\draw (1.1,0) -- (1.5,0) node[below] {$s_1$};
	\draw (0,0.5) arc (90:0:0.5);
	\draw (-0.5,0) arc (180:90:0.5) node[draw,rectangle,fill=white,inner sep=0pt] {\begin{tikzpicture}[line width=1pt]
	\draw (-0.1,-0.1) -- (0.1,0.1); 
	\draw (-0.1,0.1) -- (0.1,-0.1); 
	\end{tikzpicture}};
	\end{scope}

\end{tikzpicture}
\end{equation}
The momentum factor will be the same for all three diagrams while the time factors on the other hand are determined by integrations from the initial coupling to the power spectrum at time $t=0$ to either $s_1$ or $s_2$ giving the sum
\begin{align}
	\frac{1}{2}\lp e^{s_2} - 1\rp^2 - \lp e^{s_2} - 1\rp\lp e^{s_1} - 1\rp + \frac{1}{2}\lp e^{s_1} - 1\rp^2 = \frac{1}{2}\lp e^{s_2} - e^{s_1}\rp^2
\end{align}
Similarly any loop connected to the power spectrum segment with just one leg can be attached on either side of the box, but the sum of the two diagrams give a time factor of $\lp e^{s_2} - e^{s_1}\rp$ times the common time factor associated with the other leg and the common momentum factor. These considerations show that even though the segments with a power spectrum contribute with more loop diagrams than a propagator segment the extra contributions add up to the same result as a propagator segment will give. Thus we can treat any appearance of the power spectrum in the tree level expressions in the same way as a regular two--point propagator when we sum over all one--loop corrections on the right hand side of the RG equations. The two tree level diagrams for $V^{(1,2)}$ in figure \ref{fig:Vtree} can then be treated completely analogous to the $V^{(2,1)}$ calculation above. In general any tree level diagram for the multi--point propagator $V^{(n,1+m)}$ will be analogous to a tree level diagram for $V^{(n+m,1)}$ so we will now restrict our attention to the $n$--point propagators.

Each tree level diagram for the $(n+1)$--point propagator $V^{(n,1)}$ will contain $n-1$ mergings of legs on the right hand side of the diagram (see figure \ref{fig:V51ex} for an example). The number of two--point propagators in such a diagram is $2n-1$ and the number of possible one--loop diagrams is $n(2n-1)$. We can treat all these one--loop corrections systematically by adding up diagrams with loops connected to a certain segment of the tree level diagram separately.

\begin{figure}
\begin{center}
\begin{tikzpicture}[line width=1pt]

	\draw[dashed] node[above] {$s_4$} (0,0) -- (-1.1,0);
	\draw (-1.1,0) -- (-2,0) node[above] {$\quad\;\tb{k}$};
	\draw (0,0) -- (-12:2.9) node[below,rotate=-12] {$\tb{k}_4 + \tb{k}_5$};
	\draw[dashed] (-12:2.9) -- (-12:6) node[above] {$s_3$};
	\draw (-12:6) -- (-12:6.9);
	\draw[dashed] (-12:6.9) -- (-12:8) node[above] {$\tb{k}_5\quad\;$};
	\begin{scope}[shift={(-12:6)}]
		\draw (0,0) -- (12:0.9);
		\draw[dashed] (12:0.9) -- (12:2) node[above] {$\tb{k}_4\quad\;$};
	\end{scope}
	\draw (0,0) -- (12:1.9) node[above,rotate=12] {$\tb{k}_1 + \tb{k}_2 + \tb{k}_3$};
	\draw[dashed] (12:1.9) -- (12:4) node[above] {$s_2$};
	\draw (12:4) -- (12:5.9);
	\draw[dashed] (12:5.9) -- (12:8) node[above] {$\tb{k}_1\quad\;$};
	\begin{scope}[shift={(12:4)}]
		\draw (0,0) -- (-12:0.9) node[below,rotate=-12] {$\tb{k}_2 + \tb{k}_3$};
		\draw[dashed] (-12:0.9) -- (-12:2) node[above] {$s_1$};
		\draw (-12:2) -- (-12:2.9);
		\draw[dashed] (-12:2.9) -- (-12:4) node[above] {$\tb{k}_3\quad\;$};
		\begin{scope}[shift={(-12:2)}]
			\draw (0,0) -- (12:0.9);
			\draw[dashed] (12:0.9) -- (12:2) node[above] {$\tb{k}_2\quad\;$};
		\end{scope}
	\end{scope}
	\draw[dotted] (-2,-0.5) -- (-2,0.5) node[left] {$\eta_a$};
	\draw[dotted] (-14:8.1) -- (14:8.1) node[right] {$\eta_b$};

\end{tikzpicture}
\end{center}
\caption{A tree level diagram for $V^{(5,1)}$.}
\label{fig:V51ex}
\end{figure}

If we consider a segment where two legs merge (as in the bottom part of the diagram in figure \ref{fig:V51ex}, where the legs with momenta $\tb{k}_4$ and $\tb{k}_5$ merge between the times $\eta_b$ and $s_4$) we can use the calculation for $V^{(2,1)}$ to add all the loop diagrams with both legs on the same segment. The resulting momentum and time factor will be
\begin{align}\label{eq:merge2}
	|\tb{k}_i + \tb{k}_j|^2\lp e^{s_l} - e^{\eta_b}\rp^2
\end{align}
where $s_l$ is the time at which the next merging takes place and $\tb{k}_i$ and $\tb{k}_j$ are the momenta before the merging. Similarly there are three loops with just one leg attached to the segment and the other leg on a propagator with momentum $\tb{p}$ going between times $s_{p_1}$ and $s_{p_2}$. These can be added to give the momentum and time factors
\begin{align}\label{eq:merge1}
\begin{split}
	2\tb{p}\cdot\lp\tb{k}_i\lp e^{s_m} - e^{\eta_b}\rp + \tb{k}_j\lp e^{s_m} - e^{\eta_b}\rp + \lp\tb{k}_i + \tb{k}_j\rp\lp e^{s_l} - e^{s_m}\rp\rp\lp e^{s_{p_2}} - e^{s_{p_1}}\rp \\
	= 2\tb{p}\cdot\lp\tb{k}_i + \tb{k}_j\rp\lp e^{s_l} - e^{\eta_b}\rp\lp e^{s_{p_2}} - e^{s_{p_1}}\rp
\end{split}
\end{align}
where we see that the time of merging, $s_m$, does not appear in the final result. This taken together with equation \eqref{eq:merge2} shows that the segment can be treated as just one two--point propagator with momentum $\tb{k}_i + \tb{k}_j$ between the times $\eta_b$ and $s_l$ when we sum over one--loop diagrams.

Using this approach iteratively for every merging of legs in the tree level expression under consideration we end up with just two legs merging at the final vertex where the result will depend only on the total momentum $\tb{k} = -\textstyle\sum\tb{k}_i$ and the initial and final times $\eta_a$ and $\eta_b$. The intermediate times will only appear in the explicit tree level expression that will be common to all the one--loop diagrams. As argued above the approach works also for tree level diagrams with more than one outgoing leg on the left hand side where there will be one or more legs with a power spectrum. The only difference being a change of sign in the momenta considered so that $\tb{k} = \textstyle\sum\tb{p}_i = -\textstyle\sum\tb{k}_j$ where $\tb{p}_i$ are the momenta on the left hand side and $\tb{k}_j$ the momenta on the right hand side.

When the sum over all one--loop corrections is taken the resulting momentum and time factors will be independent of which particular tree level realization we are considering so adding all possible one--loop diagrams for a given multi--point propagator will give
\begin{align}\label{eq:genRGG}
	\partial_\lambda V_{a_1\cdots a_mb_1\cdots b_n,\lambda}^{(n,m)} = -k^2\frac{\lp e^{\eta_a} - e^{\eta_b}\rp^2}{2}\frac{1}{3}\int{\text{d}^3\tb{q}}\;\frac{P^0(q)}{q^2}\delta(\lambda - q)\sum\lp\text{tree level diagrams}\rp
\end{align}
which is identical to equation \eqref{eq:V21lkdif}. Promoting the tree level expression on the right hand side to the full $\lambda$--dependent one before integrating, we get the general large--$k$ result
\begin{align}
	V_{a_1\cdots a_mb_1\cdots b_n,\lambda\rightarrow\infty}^{(n,m)} = V_{a_1\cdots a_mb_1\cdots b_n,\text{tree}}^{(n,m)}\exp\lp-\frac{k^2\sigma_v^2}{2}\lp e^{\eta_a} - e^{\eta_b}\rp^2\rp
\end{align}
This result has been obtained for $m=1$ and $\eta_b=0$ in \cite{BernardeauG}. The fact that it generalizes to $m>1$ is not obvious from the construction of this class of multi--point propagators. As explained in Section \ref{sec:con} they can be constructed by combining two or more $V^{(n-1,1)}$ $n$--point propagators via couplings to the initial statistics, but the result is not just a product of the $V^{(n-1,1)}$ results. There are additional couplings between the individual $n$--point propagators that modify the result.

Let us consider a specific diagram for $V^{(1,2)}$ to illustrate this point
\begin{equation}\label{eq:V12dia}
\begin{tikzpicture}[line width=1.7pt]
	\draw[dashed] (0,0) -- (190:1.8cm) node[below] {$\tb{p}_2$};
	\draw (190:1.8) -- (190:3.4);
	\draw (0,0) -- (170:0.5cm);
	\draw[dashed] (170:0.5) -- (170:1.2);
	\draw (170:1.2) -- (170:1.7) node[draw,rectangle,fill=white,inner sep=1mm,rotate=-10] {} -- (170:2.2cm);
	\draw[dashed] (170:2.2cm) -- (170:2.9) node[above] {$\tb{p}_1$};
	\draw (170:2.9) -- (170:3.4);
	\begin{scope} [shift={(170:1.2cm)},line width=1pt]
		\draw[dotted] (260:0.3cm) -- (80:0.4cm) node[above] {$\eta_b$};
	\end{scope}
	\begin{scope} [shift={(170:2.2cm)},line width=1pt]
		\draw[dotted] (260:0.4cm) -- (80:0.4cm) node[above] {$\eta_b$};
	\end{scope}
	\draw (0,0) -- (0.9,0) node[above] {$-\tb{k}$};
	\draw[dashed] (0.9,0) -- (1.9,0);
	\draw[line width=1pt,dotted] (193:3.45) -- (167:3.45) node[left] {$\eta_a$};
	\draw[line width=1pt,dotted] (1.9,-0.5) -- (1.9,0.5) node[right] {$\eta_b$};
	\fill[white] (0,0) circle (1mm);
  \fill[gray,path fading=east,fading transform={rotate=-45}] (0,0) circle (1mm);  
  \draw (0,0) circle (1mm);
\end{tikzpicture}
\end{equation}
Here we have a two--point propagator $V_{ab}^{(1,1)} = G_{ab}$ running from $\eta_b$ to $\eta_a$ with momentum $\tb{p}_1$, and a three--point propagator $V_{def}^{(2,1)}$ running from $\eta_b$ to $\eta_a$ merging the two momenta $\tb{p}_1$ and $-\tb{k}$ into $\tb{p}_2$. The large--$k$ results for these two have the damping factors
\begin{align}
	\exp\lp-\frac{p_1^2\sigma_v^2}{2}\lp e^{\eta_a} - e^{\eta_b}\rp^2\rp \quad \text{and} \quad \exp\lp-\frac{p_2^2\sigma_v^2}{2}\lp e^{\eta_a} - e^{\eta_b}\rp^2\rp
\end{align}
respectively while the combination in the diagram in equation \eqref{eq:V12dia} is damped by
\begin{align}
	\exp\lp-\frac{k^2\sigma_v^2}{2}\lp e^{\eta_a} - e^{\eta_b}\rp^2\rp
\end{align}
where $\tb{k} = \tb{p}_1 + \tb{p}_2$. The additional couplings that arise in the combination of the two $n$--point propagators correspond to the term $2\tb{p}_1\cdot\tb{p}_2$ in $k^2$.

This is analogous to bremsstrahlung processes in particle physics where couplings to an external potential leads to emissions or absorptions of particles. Here the initial conditions take the role of the external potential while each Fourier mode acts as an individual particle and the coupling to the initial conditions induces couplings between the two modes.

\subsection{Large--$k$ limit with Non--Gaussian Initial Conditions}\label{sec:largekNG}

The large--$k$ assumption of Section \ref{sec:largekG} is equivalent to assuming that the momenta $\tb{k}_i$ running along the legs of the tree level diagrams are much greater than the momentum $\tb{q}$ in the power spectrum integration kernel defined in Section \ref{sec:RGpropG}. We will also use this assumption in the case of initial non--Gaussianities so that the momenta $\tb{k}_i$ are always much greater than the loop momenta $\tb{q}_j$ in the integration kernels of equation \eqref{eq:kern}.

As seen in equation \eqref{eq:RGW} the right hand side of the RG equations can be split into separate contributions from each order of the initial statistics. The contribution from the power spectrum will not change due to the presence of initial non--Gaussianities so we can reuse the results from Section \ref{sec:largekG} directly. If we do not rewrite the momentum factors as in equation \eqref{eq:qint} the end result in equation \eqref{eq:genRGG} can be expressed as
\begin{align}
	\partial_\lambda V_{a_1\cdots a_mb_1\cdots b_n,\lambda}^{(n,m)} = -V_{a_1\cdots a_mb_1\cdots b_n,\lambda}^{(n,m)}\frac{\lp e^{\eta_a} - e^{\eta_b}\rp^2}{2}\int{\text{d}^3\tb{q}}\;\frac{\lp\tb{k}\cdot\tb{q}\rp^2}{q^4}P^0(q)\delta(\lambda - q)
\end{align}
where we have promoted the tree level expression to the full $\lambda$--dependent one.

Next we turn to the contribution from the bispectrum. As we argued in Section \ref{sec:RGpropNG} we will only consider two--loop diagrams as the ones in figure \ref{fig:biRGG}. Out of these only the three in the first line will contribute at tree level, and the first one will have three vertices coupling directly to the large momentum in the tree level diagram as opposed to just two for the last two diagrams. We can see from the vertex coefficients $\alpha$ and $\beta$ in equation \eqref{eq:alfbet} that these couplings introduce a hierarchy between the diagrams of the order
\begin{align}
	\frac{\tb{k}\cdot\tb{q}_1}{q_1^2}\frac{\tb{k}\cdot\tb{q}_2}{q_2^2}\frac{\tb{k}\cdot\tb{q}_3}{q_3^3} \gg \frac{\tb{k}\cdot\tb{q}_1}{q_1^2}\frac{\tb{k}\cdot\lp\tb{q}_2 + \tb{q}_3\rp}{|\tb{q}_2 + \tb{q}_3|^2}\frac{\tb{q}_2\cdot\tb{q}_3}{q_2^2}
\end{align}
so in the large--$k$ regime we need only consider the first kind of diagrams where all legs are connected directly to the tree level diagram. This argument applies equally well to the case of higher order statistics.

Similar to equation \eqref{eq:Pkerntree} the bispectrum integration kernel reduces to 
\begin{equation}
\begin{tikzpicture}[line width=1.7pt]

	\foreach \angle in {30,-30} {\draw (0,0) -- (\angle:0.8);}
	\draw (0,0) node[draw,rectangle,fill=white,inner sep=0,line width=1.5pt] {\begin{tikzpicture} [line width=1.5pt]
	\draw (-0.1,-0.1) -- (0.1,0.1); 
	\draw (-0.1,0.1) -- (0.1,-0.1); 
	\end{tikzpicture}} -- (-0.8,0);
	\path (0.7,0) -- (0.8,0) node[right] {$\approx u_au_bu_c\theta_\text{H}(s_1)\theta_\text{H}(s_2)\theta_\text{H}(s_3)B^0(\tb{q}_1,\tb{q}_2,\tb{q}_3)\delta\lp\lambda - \textstyle\sum q_i\rp$};

\end{tikzpicture}
\end{equation}
when the initial fields are in the growing mode. This means that all three vertices that attach the bispectrum to the tree level diagram can be approximated using equation \eqref{eq:vertred} and the delta function in the indices ensures that we are again left with just the tree level expression times a momentum factor and a time factor as in equation \eqref{eq:1llk}. The momentum factor will be
\begin{align}
	\int{\text{d}^3\tb{q}_1\text{d}^3\tb{q}_2\text{d}^3\tb{q}_3}\;\frac{\tb{k}_1\cdot\tb{q}_1}{q_1^2}\frac{\tb{k}_2\cdot\tb{q}_2}{q_2^2}\frac{\tb{k}_3\cdot\tb{q}_3}{q_3^2}B^0(\tb{q}_1,\tb{q}_2,\tb{q}_3)\delta\lp\lambda - \textstyle\sum q_i\rp
\end{align}
where $\tb{k}_i$ are the three momenta running along the segments of the tree level diagram where the loop is attached. 

The time factor consists of the integral $\textstyle\int{\text{d}s_1\text{d}s_2\text{d}s_3}\;e^{s_1+s_2+s_3}$ where $s_i$ are the times of interaction with the loop momenta and the limits of integration depends on which segments the loops are attached to.

These observations generalize readily to higher order statistics, so the task is now to consider how these momentum and time factors add when we sum over all diagrams on the right hand side of the RG equations. We start by considering a segment with the merging of just two legs in the tree level diagram which is equivalent to the three--point propagator $V^{(2,1)}$.

\subsubsection{Large-$k$ limit of $V^{(2,1)}$, the non--Gaussian case}

For generality we will consider the $n$th order of the initial statistics. This means that there will be $n$ loop legs that should be attached to the tree level diagram
\begin{equation}\label{eq:treeV21dia}
\begin{tikzpicture}[line width=1pt]
	\draw[dashed] node[above] {$s_v$} (0,0) -- (-0.8,0) node[above] {$\tb{k}$};
	\draw (-0.8,0) -- (-1.5,0);
	\draw (0,0) -- (20:0.7) node[above] {$\tb{k}_1$};
	\draw[dashed] (20:0.7) -- (20:1.5);
	\draw (0,0) -- (-20:0.7) node[below] {$\tb{k}_2$};
	\draw[dashed] (-20:0.7) -- (-20:1.5);
	\draw[dotted] (-1.5,-0.5) -- (-1.5,0.5) node[left] {$\eta_a$};
	\draw[dotted] (-23:1.5) -- (23:1.5) node[right] {$\eta_b$};
\end{tikzpicture}
\end{equation}
We will denote the momenta in the loop legs as $\tb{q}_1\cdots\tb{q}_n$ and the times when they interact as $s_1\cdots s_n$. It is not necessary to distinguish between the different loop legs since we integrate over both $\tb{q}_i$ and $s_i$ and the initial statistics are symmetrized in their momentum dependencies.

If there are $m$ loop legs attached to a specific tree level leg in between times $s_{v_1}$ and $s_{v_2}$ the time factor from that leg is determined by the $m$ integrals
\begin{align}
	\int^{s_{v_2}}_{s_{v_1}}\text{d}s_m\;e^{s_m} \int^{s_m}_{s_{v_1}}\text{d}s_{m-1}\;e^{s_{m-1}}\cdots \int^{s_2}_{s_{v_1}}\text{d}s_1\;e^{s_1}
\end{align}
which by induction can be shown to yield
\begin{align}\label{eq:timeind}
	\int^{s_{v_2}}_{s_{v_1}}\text{d}s_m\;e^{s_m}\frac{1}{(m-1)!}\lp e^{s_m} - e^{s_{v_1}}\rp^{(m-1)} = \frac{1}{m!}\lp e^{s_{v_2}} - e^{s_{v_1}}\rp^m
\end{align}
Thus if we have $m$ loop legs attached to the left leg in equation \eqref{eq:treeV21dia} they will contribute with a time and momentum factor given by
\begin{align}\label{eq:mlegsl}
	\frac{1}{m!}\lp e^{\eta_a} - e^{s_v}\rp^m\prod_{i=1}^m\frac{\tb{k}\cdot\tb{q}_i}{q_i^2}
\end{align}

The remaining $n-m$ legs should be attached to the two legs on the right hand side in equation \eqref{eq:treeV21dia}. The resulting $n-m+1$ diagrams will have a common time factor of $\lp e^{s_v} - e^{\eta_b}\rp^{n-m}$, but the fraction connected to the time factor (as in equation \eqref{eq:timeind}) and the momentum factor will depend on exactly how many loop legs are connected to which tree level leg. With just one leg to attach we get two terms
\begin{align}
-\tb{k}_1\cdot\tb{q}_i - \tb{k}_2\cdot\tb{q}_i = \tb{k}\cdot\tb{q}_i
\end{align}
where we have left out the factor of $1/q_i^2$. With two legs one of the three terms can be rewritten using the invariance under $\tb{q}_i \leftrightarrow \tb{q}_j$ so that
\begin{align}
\begin{split}
	\frac{1}{2}&(\tb{k}_1\cdot\tb{q}_i)(\tb{k}_1\cdot\tb{q}_j) + (\tb{k}_1\cdot\tb{q}_i)(\tb{k}_2\cdot\tb{q}_j) + \frac{1}{2}(\tb{k}_2\cdot\tb{q}_i)(\tb{k}_2\cdot\tb{q}_j) \\ 
	&= \frac{1}{2}\lp(\tb{k}_1\cdot\tb{q}_i)(\tb{k}_1\cdot\tb{q}_j) + (\tb{k}_1\cdot\tb{q}_i)(\tb{k}_2\cdot\tb{q}_j) + (\tb{k}_2\cdot\tb{q}_i)(\tb{k}_1\cdot\tb{q}_j) + (\tb{k}_2\cdot\tb{q}_i)(\tb{k}_2\cdot\tb{q}_j)\rp \\ 
	&= \frac{1}{2}(\tb{k}\cdot\tb{q}_i)(\tb{k}\cdot\tb{q}_j)
\end{split}
\end{align}
This result generalizes easily for $n-m$ legs giving
\begin{align}\label{eq:mlegsr}
	\frac{1}{(n-m)!}\lp e^{s_v} - e^{\eta_b}\rp^{(n-m)}\prod_{i=m}^n\frac{\tb{k}\cdot\tb{q}_i}{q_i^2}
\end{align}
where we have reinstated in the factors of $1/q_i^2$ and the common time factor.

Equation \eqref{eq:mlegsr} taken together with equation \eqref{eq:mlegsl} gives the total momentum and time factor obtained by adding all diagrams with $m$ loop legs attached to the left leg in equation \eqref{eq:treeV21dia}. Now adding all possible loop diagrams from the $n$th order initial statistic results in a sum over these factors from $m=0$ to $m=n$
\begin{align}
\begin{split}
	&\sum_{m=0}^n\frac{1}{m!}\frac{1}{(n-m)!}\lp e^{\eta_a} - e^{s_v}\rp^{m}\lp e^{s_v} - e^{\eta_b}\rp^{(n-m)} \prod_{i=1}^n\frac{\tb{k}\cdot\tb{q}_i}{q_i^2} \\
	&=\frac{1}{n!}\prod_{i=1}^n\frac{\tb{k}\cdot\tb{q}_i}{q_i^2} \sum_{m=0}^n\binom{n}{m}\lp e^{\eta_a} - e^{s_v}\rp^{m}\lp e^{s_v} - e^{\eta_b}\rp^{(n-m)} = \frac{1}{n!}\lp e^{\eta_a} - e^{\eta_b}\rp^{n}\prod_{i=1}^n\frac{\tb{k}\cdot\tb{q}_i}{q_i^2}
\end{split}
\end{align}
where we see that just as in the Gaussian case we have eliminated the dependence on the intermediate time $s_v$ so that it will only appear in the explicit expression for the tree level diagram in equation \eqref{eq:treeV21dia}.

The full result for the RG equation for $V^{(2,1)}$ after promoting the tree level expression to the full non--linear one is then
\begin{align}\label{eq:NGRGV21}
\begin{split}
	\partial_\lambda V^{(2,1)}_{abc,\lambda} =& -V^{(2,1)}_{abc,\lambda}\Bigg(\frac{\lp e^{\eta_a} - e^{\eta_b}\rp^2}{2!}\int{\text{d}^3\tb{q}}\;\frac{\lp\tb{k}\cdot\tb{q}\rp^2}{q^4}P^0(q)\delta(\lambda - q) \\
	&- \frac{\lp e^{\eta_a} - e^{\eta_b}\rp^3}{3!}\int{\text{d}^3\tb{q}_{1,2,3}}\;\prod_{i=1}^3\frac{\tb{k}\cdot\tb{q}_i}{q_i^2}B^0(\tb{q}_1,\tb{q}_2,\tb{q}_3)\delta\lp\lambda - \textstyle\sum q_i\rp - \cdots\Bigg)
	\end{split}
\end{align}
We can express the initial statistics in terms of correlation functions of the initial density perturbation $\delta_0$
\begin{align}
\begin{split}
	\delta(\tb{q}_1 + \tb{q}_2)P^0(q) &= \left\langle\delta_0(\tb{q}_1)\delta_0(\tb{q}_2)\right\rangle \\
	\delta(\tb{q}_1 + \tb{q}_2 + \tb{q}_3)B^0(\tb{q}_1,\tb{q}_2,\tb{q}_3) &= \left\langle\delta_0(\tb{q}_1)\delta_0(\tb{q}_2)\delta_0(\tb{q}_3)\right\rangle \\
	&\;\;\vdots
\end{split}
\end{align}
This enables us to rewrite the parentheses in equation \eqref{eq:NGRGV21} more compactly so that the end result after integrating and letting $\lambda \rightarrow \infty$ is
\begin{align}\label{eq:V21resNG}
	V^{(2,1)}_{abc,\lambda\rightarrow\infty} = V^{(2,1)}_{abc,\text{tree}}\exp\lp\sum_{n=2}^\infty\frac{\lp e^{\eta_a} - e^{\eta_b}\rp^{n}}{n!}\left\langle\prod_{i=1}^n\int{\text{d}^3\tb{q}_i}\;\frac{\tb{k}\cdot\tb{q}_i}{q_i^2}\delta_0(\tb{q}_i)\right\rangle\rp
\end{align}
The change in sign for the power spectrum term compared to equation \eqref{eq:NGRGV21} is due to the fact that the delta function $\delta\lp\tb{q}_1 + \tb{q}_2\rp$ has not yet been enforced in this expression.

\subsubsection{Large--$k$ limit of general multi--point propagators, the non--Gaussian case}

To generalize to all multi--point propagators we need to go through the same considerations as in Section \ref{sec:largekG}. For the multi--point propagators with just one outgoing leg on the left hand side we can again use the calculation for $V^{(2,1)}$ above to show that every segment that merges two legs from the right can be treated as just one two--point propagator with the sum of the two initial momenta running along it. An iterative use of this gives the general result analogous to equation \eqref{eq:V21resNG}.

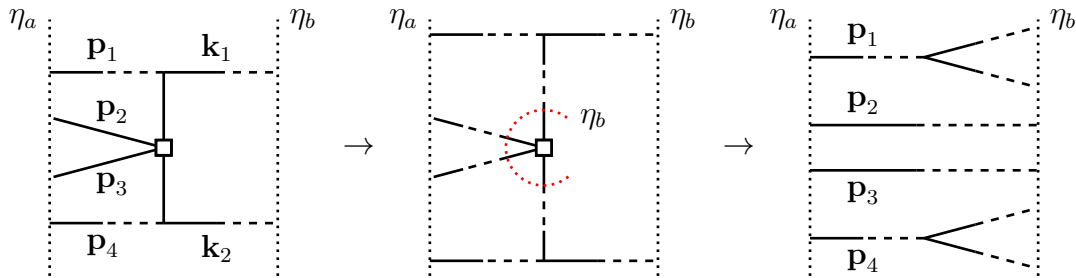
\begin{figure}
\begin{center}
\begin{tikzpicture}[line width=1pt]

	\draw (0,0) -- (165:0.7) node[above] {$\tb{p}_2$} -- (165:1.5);
	\draw (0,0) node[draw,rectangle,fill=white,inner sep=1mm] {} -- (195:0.7) node[below] {$\tb{p}_3$} -- (195:1.5);
	\foreach \angle in {90,-90} {\draw (0,0) node[draw,rectangle,fill=white,inner sep=1mm] {} -- (\angle:1);}
	\draw[dashed] (90:1) -- (-0.8,1) node[above] {$\tb{p}_1$};
	\draw (-0.8,1) -- (-1.5,1);
	\draw (90:1) -- (0.7,1) node[above] {$\tb{k}_1$};
	\draw[dashed] (0.7,1) -- (1.5,1);
	\draw[dashed] (-90:1) -- (-0.8,-1) node[below] {$\tb{p}_4$};
	\draw (-0.8,-1) -- (-1.5,-1);
	\draw (-90:1) -- (0.7,-1) node[below] {$\tb{k}_2$};
	\draw[dashed] (0.7,-1) -- (1.5,-1);
	\draw[dotted] (-1.5,-1.7) -- (-1.5,1.7) node[left] {$\eta_a$};
	\draw[dotted] (1.5,-1.7) -- (1.5,1.7) node[right] {$\eta_b$};
	\path (1.5,0) -- (2.2,0) node[right] {$\rightarrow$};

	\begin{scope}[xshift=5cm]
	\foreach \angle in {195,165,90,-90} {\draw (0,0) node[draw,rectangle,fill=white,inner sep=1mm] {} -- (\angle:0.5); \draw[dashed] (\angle:0.5) -- (\angle:1.1); \draw (\angle:1.1) -- (\angle:1.5);}
	\draw[dashed] (90:1.5) -- (-0.8,1.5);
	\draw (-0.8,1.5) -- (-1.5,1.5);
	\draw (90:1.5) -- (0.7,1.5);
	\draw[dashed] (0.7,1.5) -- (1.5,1.5);
	\draw[dashed] (-90:1.5) -- (-0.8,-1.5);
	\draw (-0.8,-1.5) -- (-1.5,-1.5);
	\draw (-90:1.5) -- (0.7,-1.5);
	\draw[dashed] (0.7,-1.5) -- (1.5,-1.5);
	\draw[dotted] (-1.5,-1.7) -- (-1.5,1.7) node[left] {$\eta_a$};
	\draw[dotted] (1.5,-1.7) -- (1.5,1.7) node[right] {$\eta_b$};
	\draw[dotted,red] (310:0.5) arc (310:50:0.5) node[right,black] {$\eta_b$};
	\path (1.5,0) -- (2.2,0) node[right] {$\rightarrow$};
	\end{scope}
	
	\begin{scope}[xshift=10cm]
		\begin{scope}[yshift=1.2cm]
		\draw[dashed] (0,0) -- (-0.8,0) node[above] {$\tb{p}_1$};
		\draw (-0.8,0) -- (-1.5,0);
		\foreach \angle in {15,-15} {\draw (0,0) -- (\angle:0.7); \draw[dashed] (\angle:0.7) -- (\angle:1.5);}
		\end{scope}
		\begin{scope}[yshift=0.3cm]
		\draw (-0.1,0) -- (-0.8,0) node[above] {$\tb{p}_2$} -- (-1.5,0);
		\draw[dashed] (-0.1,0) -- (1.5,0);
		\end{scope}
		\begin{scope}[yshift=-0.3cm]
		\draw (-0.1,0) -- (-0.8,0) node[below] {$\tb{p}_3$} -- (-1.5,0);
		\draw[dashed] (-0.1,0) -- (1.5,0);
		\end{scope}
		\begin{scope}[yshift=-1.2cm]
		\draw[dashed] (0,0) -- (-0.8,0) node[below] {$\tb{p}_4$};
		\draw (-0.8,0) -- (-1.5,0);
		\foreach \angle in {15,-15} {\draw (0,0) -- (\angle:0.7); \draw[dashed] (\angle:0.7) -- (\angle:1.5);}
		\end{scope}
	\draw[dotted] (-1.5,-1.7) -- (-1.5,1.7) node[left] {$\eta_a$};
	\draw[dotted] (1.5,-1.7) -- (1.5,1.7) node[right] {$\eta_b$};
	\end{scope}

\end{tikzpicture}
\end{center}
\caption{Example of a tree level diagram with a trispectrum merging $4$ legs. The reasoning behind the last two diagrams is explained in the text.}
\label{fig:statmerge}
\end{figure}

On the other hand the case of multi--point propagators with more than one outgoing leg on the left hand side is more complicated in the presence of initial non--Gaussianities. Apart from tree level segments that include an initial power spectrum there will also be segments that merge two or more legs by direct couplings to higher order initial statistics (see the first diagram in figure \ref{fig:statmerge} for an example). We can split such diagrams into two parts by letting all the legs from the initial statistic go out to the time $\eta_b$ before proceeding to the next interaction as seen in the second diagram of figure \ref{fig:statmerge}. Now any loop correction to such a tree level diagram that has a leg attached inside the center circle will not contribute to the RG equation when we sum over all diagrams, because the sum of the momenta coming from the initial statistic has to be zero. It is equivalent to the arguments leading up to equation \eqref{eq:mlegsr} except with $\tb{k}=0$. This means that we can treat the diagram as separate multi--point propagators all going from $\eta_b$ to $\eta_a$ as shown in the last part of figure \ref{fig:statmerge}.

Each of these separate sections can again be reduced to a single two--point propagator carrying momentum $\tb{p}_i$ and calculating loop diagrams along the same lines as was done for $V^{(2,1)}$ we arrive at the same final result as seen in equation \eqref{eq:V21resNG}
\begin{align}\label{eq:VresNG}
	V^{(n,m)}_{a_1\cdots a_mb_1\cdots b_n,\lambda\rightarrow\infty} = V^{(n,m)}_{a_1\cdots a_mb_1\cdots b_n,\text{tree}}\exp\lp\sum_{n=2}^\infty\frac{\lp e^{\eta_a} - e^{\eta_b}\rp^{n}}{n!}\left\langle\prod_{i=1}^n\int{\text{d}^3\tb{q}_i}\;\frac{\tb{k}\cdot\tb{q}_i}{q_i^2}\delta_0(\tb{q}_i)\right\rangle\rp
\end{align}
This is the result obtained in \cite{Bernardeau} for the $n$--point propagators $V^{(n-1,1)}$, that we have shown also holds for the generalization $V^{(n,m)}$ with $m>1$.

The multi--point propagators can be interpreted as a measure of how well the initial conditions are preserved later in time. The linear two--point propagator for instance tells us that the linear evolution preserves the initial statistics perfectly because it is normalized to $1$ at all times and scales. The full propagators on the other hand all show an exponential damping at large $k$--values meaning that the memory of the initial statistics is erased at small scales where the non--linear collapse of the matter density peaks has proceeded far beyond the regime where standard perturbation theory is valid.

The generalized results presented here show that even though the multi--point propagators $V^{(n,m)}$ with $m>1$ are in a sense more connected to the initial conditions through their tree level expressions, they lose memory of the statistics at the same rate as the $n$--point propagators.

It was noted in \cite{Bernardeau} that in an isotropic background universe the terms in the exponential of equation \eqref{eq:VresNG} cannot depend on the direction of $\tb{k}$. We saw in the Gaussian case that the contribution from the power spectrum is proportional to $k^2$. Considering the term from the bispectrum the $\tb{k}$--dependence is contained in the factor
\begin{align}
	f(k) = \int{\text{d}^3\tb{q}_1\text{d}^3\tb{q}_2\text{d}^3\tb{q}_3}\;\frac{\tb{k}\cdot\tb{q}_1}{q_1^2}\frac{\tb{k}\cdot\tb{q}_2}{q_2^2}\frac{\tb{k}\cdot\tb{q}_3}{q_3^2}B^0(\tb{q}_1,\tb{q}_2,\tb{q}_3)
\end{align}
We can perform an explicit integration over the orientation of \tb{k} yielding $4\pi f(k) = 0$, i.e. the bispectrum contribution vanishes. This will be the case for all the terms with uneven $n$ in the sum. Only the even terms survive and the first correction to the Gaussian case comes from the trispectrum. 

\section{Conclusions}\label{chap:conc}

In this paper we have investigated the large--$k$ limit of multi--point propagators in the framework of the RG formalism presented in \cite{Matarrese}. We have reproduced the results of \cite{BernardeauG} in the case of Gaussian initial conditions and \cite{Bernardeau} in the presence of initial non--Gaussianities for the special class of $n$--point propagators that was introduced in \cite{BernardeauG}. In addition to this we have obtained large--$k$ results for the generalized multi--point propagators $V^{(n,m)}$ with $m>1$ that play a role in the RG formalism along side the $n$--point propagators.

All the multi--point propagators show the same exponential damping at large $k$--values that was first found for the two--point propagator in \cite{Crocce} for the Gaussian case. When initial non--Gaussianities are considered the first non--zero correction comes from the trispectrum which in most cases will only give minor deviations from the Gaussian result. In \cite{Bernardeau} a specific local model was considered with $f_\text{NL} = 10^3$ and $g_\text{NL} = 10^7$ that gave up to $2\%$ weaker decay at $z=0$. With more realistic values for $f_\text{NL}$ and $g_\text{NL}$ the deviations are much smaller.

In \cite{BernardeauG} and \cite{Bernardeau} the large--$k$ limit of the $n$--point propagators is used directly to construct the power spectrum and bispectrum at late times and on small scales. The result is then combined with standard one--loop perturbation theory on large scales by interpolating between the two regimes. In the RG approach the large--$k$ limit of multi--point propagators could be used in a similar way as it has been for the two--point propagator in \cite{Matarrese} where the time dependence from the large--$k$ result is assumed also to be valid for small $k$--values. The large--$k$ results for the multi--point propagators represent a renormalization of the 1PI vertices so an implementation of the RG equations that go beyond the tree level approximation for the vertices might be possible.

Another possibility would be to calculate the power spectrum and bispectrum in the presence of initial non--Gaussianities. The framework to do this has been set up in Section \ref{chap:largek}, but the RG equations become more complicated when we are also interested in the behaviour at small $k$--values. In this regime the kinds of diagrams in equations \eqref{eq:treeRGcon} and \eqref{eq:bione} that could be neglected in Section \ref{chap:largek} play an important role. In particular the power spectrum will get a new one--loop contribution from the initial bispectrum given by the diagram
\begin{equation}
\begin{tikzpicture}[line width=1.7pt]
	
	\draw[dashed] (0,0) -- (-0.8,0);
	\draw (-0.8,0) -- (-1.4,0);
	\draw (0,0) arc (180:0:0.3);
	\draw (0,0) arc (-180:0:0.3);
	\draw (1.4,0) -- (0.6,0) node[draw,rectangle,fill=white,inner sep=0pt,line width=1.5pt] {\begin{tikzpicture}[line width=1.5pt]
	\draw (-0.1,-0.1) -- (0.1,0.1); 
	\draw (-0.1,0.1) -- (0.1,-0.1); 
	\end{tikzpicture}};
	\fill[white] (0,0) circle (1mm);
  \fill[gray,path fading=east,fading transform={rotate=-45}] (0,0) circle (1mm);  
  \draw (0,0) circle (1mm);
	
\end{tikzpicture}
\end{equation}
It seems that including a diagram like this in the numerical solution of the RG equations should be possible. When it comes to the bispectrum the multi--point propagator $V^{(1,2)}$ provides us with the diagram
\begin{equation}
\begin{tikzpicture}[line width=1.7pt]
	
	\foreach \angle in {150,210} {\draw[dashed] (0,0) -- (\angle:0.7); \draw (\angle:0.7) -- (\angle:1.2);}
	\draw (0,0) -- (0.6,0) node[draw,rectangle,fill=white,inner sep=0pt,line width=1.5pt] {\begin{tikzpicture}[line width=1.5pt]
	\draw (-0.1,-0.1) -- (0.1,0.1); 
	\draw (-0.1,0.1) -- (0.1,-0.1); 
	\end{tikzpicture}} -- (1.2,0);
	\fill[white] (0,0) circle (1mm);
  \fill[gray,path fading=east,fading transform={rotate=-45}] (0,0) circle (1mm);  
  \draw (0,0) circle (1mm);
	
\end{tikzpicture}
\end{equation}
that might work as a first approximation to the renormalized bispectrum.

The approach followed in the calculations presented in this article corresponds to resumming only loop diagrams where all the loop legs are attached directly to the multi--point propagator in standard perturbation theory. In \cite{Anselmi} it was shown that in the Gaussian case it is possible to resum another class of diagrams in the large--$k$ limit by essentially using a resummed power spectrum in the integration kernel of equation \eqref{eq:Pkerntree}. It would be interesting to see if their result can be reproduced in the case of multi--point propagators.

\section*{Acknowledgements}

I thank M. Pietroni and S. Hannestad for useful discussions and comments.

\end{document}